\documentclass[reqno,11pt]{amsart}
\usepackage{graphicx}
\usepackage{amssymb}
\usepackage{slashed}
\usepackage[mathscr]{eucal}
\numberwithin{equation}{section}
\allowdisplaybreaks[1]

\newcommand{\SetFigFont}[3]{}

\title[The Fermionic Signature Operator in Rindler Space-Time]
{The Fermionic Signature Operator and \\ Quantum States in Rindler Space-Time}
\author[F.\ Finster, S.\ Murro]{Felix Finster, Simone Murro}
\author[C.\ R\"oken]{Christian R\"oken \\ \\ June 2016}
\thanks{F.F.\ and C.R.\ are supported by the DFG research grant ``Dirac Waves in the Kerr Geometry: Integral Representations, Mass Oscillation Property and the Hawking Effect.''
S.M. is supported within the DFG research training group GRK 1692 ``Curvature, Cycles, and Cohomology.'' }
\address{Fakult\"at f\"ur Mathematik \\ Universit\"at Regensburg \\ D-93040 Regensburg \\ Germany}
\email{finster@ur.de, simone.murro@ur.de, christian.roeken@ur.de}

\newtheorem{Def}{Definition}[section]
\newtheorem{Thm}[Def]{Theorem}
\newtheorem{Prp}[Def]{Proposition}
\newtheorem{Lemma}[Def]{Lemma}

\newtheorem{Corollary}[Def]{Corollary}

\newcommand{\Thanks}{\vspace*{.5em} \noindent \thanks}
\newcommand{\beq}{\begin{equation}}
\newcommand{\eeq}{\end{equation}}
\newcommand{\Proof}{\begin{proof}}
\newcommand{\QED}{\end{proof} \noindent}

\newcommand{\la}{\langle}
\newcommand{\ra}{\rangle}
\newcommand{\bra}{\mathopen{<}}
\newcommand{\ket}{\mathclose{>}}
\newcommand{\Sl}{\mbox{$\prec \!\!$ \nolinebreak}}
\newcommand{\Sr}{\mbox{\nolinebreak $\succ$}}

\newcommand{\C}{\mathbb{C}}
\newcommand{\R}{\mathbb{R}}
\newcommand{\1}{\mbox{\rm 1 \hspace{-1.05 em} 1}}

\newcommand{\N}{\mathbb{N}}

\newcommand{\Cisc}{C^\infty_{\text{sc}}}

\newcommand{\Dir}{{\mathcal{D}}}
\newcommand{\D}{{\mathscr{D}}}

\renewcommand{\H}{\mathscr{H}}
\renewcommand{\tilde}[1]{\widetilde{#1}}

\DeclareMathOperator{\supp}{supp}

\newcommand{\U}{\text{\rm{U}}}

\renewcommand{\tilde}[1]{\widetilde{#1}}
\renewcommand{\hat}[1]{\widehat{#1}}

\newcommand{\scrM}{\mycal M}
\newcommand{\scrR}{\mycal R}
\newcommand{\scrN}{\mycal N}
\newcommand{\Sig}{\mathscr{S}}

\newcommand{\f}{{\mathfrak{f}}}
\newcommand{\TCPT}{T_\text{\tiny{CPT}}}

\newcommand{\fiberpairing}[2]{\Sl#1 \,|\, #2\Sr}
\newcommand{\scalarproduct}[2]{(#1 | #2)}
\newcommand{\bracket}[2]{\bra#1 | #2\ket}

\DeclareFontFamily{OT1}{rsfso}{}
\DeclareFontShape{OT1}{rsfso}{m}{n}{ <-7> rsfso5 <7-10> rsfso7 <10-> rsfso10}{}
\DeclareMathAlphabet{\mycal}{OT1}{rsfso}{m}{n}

\begin{document}

\maketitle 

\begin{abstract}
The fermionic signature operator is constructed in Rindler space-time.
It is shown to be an unbounded self-adjoint operator on the Hilbert space of solutions of the massive Dirac equation.
In two-dimensional Rindler space-time, we prove that the resulting fermionic projector state coincides with
the Fulling-Rindler vacuum. Moreover, the fermionic signature operator gives a covariant construction of
general thermal states, in particular of the Unruh state.
The fermionic signature operator is shown to be well-defined in asymptotically Rindler space-times.
In four-dimensional Rindler space-time, our construction gives rise to new quantum states.
\end{abstract}

\tableofcontents

\section{Introduction}
In quantum field theory in curved space-time, the interpretation of
physical states in terms of particles and anti-particles depends on the observer.
This becomes most apparent in the well-known Unruh effect, which shows that for the usual vacuum
state in Minkowski space, a uniformly accelerated observer experiences particles and anti-particles
in a thermal state. The observer dependence of the particle interpretation is reflected
in the freedom to choose a {\em{Fock ground state}}.
Constructing a Fock space from this Fock ground state by employing the creation operators of the particles
and anti-particles, the quantum state describing the physical system
is represented by a vector of this Fock space.

Within all possible quantum states, one can distinguish a specific class of states which are
generally regarded as being physically sensible, referred to as {\em{Hadamard states}}
(for the general context see~\cite{radzikowski, sahlmann2001microlocal}).
For Hadamard states, the Wick ordering of field operators is well-defined,
making it possible to build up a perturbative quantum field theory (see for
example~\cite{fewster2013necessity, dappiaggiDirac}
or the recent text book~\cite{rejzner}).

In the present paper, we focus on a distinguished fermionic Fock ground state, referred to as the
{\em{fermionic projector (FP) state}} (for somewhat related constructions for bosonic fields see~\cite{johnston,
sorkin2}). This state is closely linked to the {\em{fermionic signature operator}} introduced
in~\cite{finite, infinite} (based on earlier perturbative constructions in~\cite{sea}).
It cannot be associated to a local observer. Instead, it depends on the global
geometry of space-time.
Nevertheless, it is a physically sensible state, provided that
it is of Hadamard form. For a more detailed account on the physical interpretation
we refer to~\cite[Section~2.1.2]{cfs} or the discussion of a scattering process in~\cite[Section~5]{sea}.
In mathematical terms, the fermionic signature operator is a symmetric operator
on the solution space of the massive Dirac equation
in globally hyperbolic space-times. It encodes geometric information~\cite{drum} and 
gives a new covariant method for obtaining Ha\-da\-mard states~\cite{hadamard}.
The abstract construction in space-times of
finite and infinite lifetime as given in~\cite{finite, infinite} opens up the research program
to explore the fermionic signature operator in various space-times and to verify if the resulting
FP states are Hadamard.
So far, the fermionic signature operator has been studied in the examples 
of closed FRW space-times~\cite{finite},
ultrastatic space-times of infinite lifetime and de Sitter space-time~\cite{infinite},
ultrastatic slab space-times~\cite{fewster+lang}
as well as for an external potential in Minkowski space which decays at infinity~\cite{hadamard}.
Moreover, as the first example involving a potential which is neither static nor decays for large times,
in~\cite{planewave} the coupling to a plane electromagnetic wave is considered.
Finally, in~\cite{drum} various two-dimensional examples are analyzed.
In the example of a space-time slab, it is shown in~\cite{fewster+lang} that the resulting FP state
is in general not of Hadamard form, but that one gets Hadamard states if the construction is ``softened''
by a smooth cutoff function in time. On the other hand, in ultrastatic space-times of infinite
lifetime~\cite[Section~5]{infinite} as well as for
an external potential in Minkowski space (see~\cite[Theorem~1.3]{hadamard} and~\cite{planewave}),
the resulting FP state
is proven to be of Hadamard form. These results lead to the conjecture that the FP state should be of
Hadamard form provided that space-time is ``sufficiently smooth'' on its boundaries and at its asymptotic ends.
In order to challenge and quantify this conjecture, one needs to construct and analyze the fermionic signature
operator in detail in different space-times.

%

As the first example involving a horizon, we here consider two-dimensional Rindler space-time
(see~\cite{rindler1966} or~\cite[Section~6.4]{wald}).
This is of physical interest in view of the Unruh effect, which is closely related to the
Hawking effect in black hole geometries (see for example~\cite{collini}).
Also, from a mathematical point of view, the example of Rindler space-time is interesting because, although
lifetime is infinite, the methods in~\cite{infinite} do not apply. The reason is that the strong mass oscillation property
does not hold due to boundary contributions on the horizon
(for boundary contributions in a more general setting see~\cite{drago+murro}).
Instead, we adapt the construction for space-times of finite lifetime in~\cite{finite},
making it possible to define the fermionic signature operator as a densely defined unbounded operator.
The construction in this paper is covariant in the sense that it does not depend on the choice of specific coordinates.
We show that the fermionic signature operator is indeed a multiple of the Dirac Hamiltonian in Rindler coordinates
(for details see Theorem~\ref{thmFRV} below).
This means that the construction of the fermionic signature operator ``detects'' the Killing symmetry of
our space-time as described by translations in Rindler time.
We thus obtain a covariant construction of the Fulling-Rindler vacuum~\cite{Fulling:1972md} and of general thermal states like the Unruh state~\cite{Unruh:1976db} (see Corollaries~\ref{corFRV1} and~\ref{corFRV2};
for a general introduction to quantum states in Rindler space-time see for example~\cite[Chapter~5]{waldqft}).

Extending the above analysis to four-dimensional Rindler space-time, we obtain states
which are indeed different from the Fulling-Rindler vacuum and general thermal states
(see Section~\ref{secfourdim}). The physical properties of these new states
are still under investigation.

It is a main advantage of our construction that it also applies in situations without Killing symmetries.
This is made clear by considering asymptotically Rindler space-times (see Theorem~\ref{thmasy}).

\section{Preliminaries}
In this section, we recall a few basic definitions, mainly using the notation and conventions in~\cite{finite}.
We restrict attention to the two-dimensional situation (for the four-dimensional setting see Section~\ref{secfourdim}).
The two-dimensional {\em{Rindler space-time}}~$(\scrR, g)$ is isometric to
the subset of two-dimensional Minkowski space
\beq \label{scrRdef}
\scrR = \big\{ (t,x) \in \R^{1,1} \;\;\;\text{with}\;\;\; |t| < x \big\}
\eeq
with the induced line element
\beq \label{mink}
ds^2 = g_{ij}\: dx^i dx^j = dt^2 - dx^2 \:.
\eeq
We let~$S\scrR = \scrR \times \C^2$ be the trivial spinor bundle.
We work in the so-called chiral representation of the Dirac matrices
\beq \label{gammarep}
\gamma^0 = \begin{pmatrix} 0 & 1 \\ 1 & 0 \end{pmatrix} \:,\qquad
\gamma^1 = \begin{pmatrix} 0 & 1 \\ -1 & 0 \end{pmatrix} \:.
\eeq
The Dirac matrices are symmetric with respect to the {\em{spin scalar product}} defined by
\beq \label{ssprod}
\Sl \psi | \phi \Sr = \la \psi, \begin{pmatrix} 0 & 1 \\ 1 & 0 \end{pmatrix} \phi \ra_{\C^2}
\eeq
(where~$\la .,. \ra_{\C^2}$ is the canonical scalar product on~$\C^2$).
The spin scalar product is an indefinite inner product of signature~$(1,1)$.
Introducing the {\em{Dirac operator}}
\beq \label{Dirop}
\Dir := i \gamma^j \partial_j \:,
\eeq
the {\em{massive Dirac equation}} reads
\beq \label{Direq}
(\Dir-m) \psi = 0 \:,
\eeq
where~$m>0$ is the rest mass (we always work in natural units~$\hbar=c=1$).

Rindler space-time is globally hyperbolic (for example, for any~$\alpha \in (-1,1)$,
the ray~$\{ (\alpha x, x) \text{ with } x > 0 \}$
is a Cauchy surface). Taking smooth and compactly supported initial data on a Cauchy surface~$\scrN$
and solving the Cauchy problem, one obtains a Dirac solution in the class~$\Cisc(\scrR, S\scrR)$
of smooth wave functions with spatially compact support. On solutions~$\psi, \phi$
in this class, one defines the (positive definite) scalar product
\beq \label{print}
(\psi | \phi) := 2 \pi \int_\scrN \Sl \psi | \slashed{\nu} \phi\Sr|_q\: d\mu_\scrN(q) \:,
\eeq
where~$\slashed{\nu} = \gamma^j \nu_j$ denotes Clifford multiplication by the future-directed unit
normal~$\nu$, and~$d\mu_\scrN$ is the volume measure of the induced Riemannian metric
on~$\scrN$ (thus for the above ray~$\scrN = \{ (\alpha x, x) \text{ with } x > 0 \}$,
the measure~$d\mu_\scrN= \sqrt{1-\alpha^2}\, dx$ is a multiple of the Lebesgue measure).
Due to current conservation, this scalar product is independent of the choice of~$\scrN$.
Forming the completion, we obtain the Hilbert space~$(\H, (.|.))$, referred to as the
{\em{solution space}} of the Dirac equation. We denote the norm on this Hilbert space
by~$\| \psi \| := \sqrt{(\psi | \psi)}$.
Another object which will be important later on is the {\em{space-time inner product}}
\beq \label{stip}
\bra \psi|\phi \ket := \int_\scrR \Sl \psi | \phi \Sr|_q\: d\mu_\scrR(q)\:.
\eeq
This inner product is not positive definite. Moreover, one should keep in mind that
the integral in~\eqref{stip} may diverge for solutions of the Dirac equation.
However, this integral is clearly well-defined for example for compactly supported wave functions.
Thus in applications, it is important to specify the class of wave functions
for which the inner product~\eqref{stip} is to be evaluated.
Furthermore, one must verify carefully that the integral in~\eqref{stip} exists
(we will come back to this point in Section~\ref{secunbound}).

The method for constructing the fermionic signature operator~$\Sig$ is as follows:
Let~$\D(\Sig)$ be a subspace of~$\H$ such that for any~$\phi \in \D(\Sig)$, the
anti-linear mapping~$\bra . | \phi \ket : \H \rightarrow \C$ given by~\eqref{stip} is
well-defined and bounded, i.e.
\[ \big| \bra \psi|\phi \ket \big| \leq C(\phi)\, \|\psi\|  \qquad \text{for all~$\psi \in \H$} \]
for a suitable constant~$C(\phi)<\infty$. Then the Fr{\'e}chet-Riesz theorem makes it possible
to represent this anti-linear mapping by a vector~$\Sig\phi$, i.e.
\beq \label{Sdef}
\bra \psi | \phi \ket = (\psi \,|\, \Sig \phi) \qquad  \text{for all~$\psi \in \H$} \:.
\eeq
Varying~$\phi \in \D(\Sig)$, we obtain a linear mapping
\[ \Sig \::\: \D(\Sig) \rightarrow \H \:, \]
referred to as the {\em{fermionic signature operator}}.
Obviously, this operator is symmetric on the Hilbert space~$\H$.
Our goal is to show that in Rindler space-time, the domain~$\D(\Sig)$ of this
operator can be chosen as a dense subset of~$\H$, and that the fermionic signature operator
has a unique self-adjoint extension.

We conclude this section with an outline of construction of the FP state (for details see~\cite[Section~6]{hadamard}).
Assume that the fermionic signature operator~$\Sig$ has been constructed as a 
self-adjoint operator on~$\H$. Then the spectral calculus allows us to form
the operator~$f(\Sig)$ for any bounded Borel function~$f : \R \rightarrow \C$.
In particular, the operator~$\chi_{(-\infty, 0)}(\Sig)$ (where~$\chi$ denotes the characteristic
function) is the projection operator onto the negative spectral subspace.
The FP state can be understood as the state of the corresponding quasi-free Dirac field
where all one-particle states in the image of the operator~$\chi_{(-\infty, 0)}(\Sig)$
are occupied. Mathematically, this state can be obtained by applying an abstract
construction due to Araki~\cite{araki1970quasifree}.
With this in mind, the main task of the subsequent analysis is to construct~$\Sig$
and to analyze its properties. This is why in our analysis we will mainly restrict
attention to the one-particle Hilbert space~$\H$.

\section{Embedding in Minkowski Space}
For the subsequent analysis, it is often useful to regard Rindler space-time as a
subset of Minkowski space, and to embed the solution space in Rindler space-time
into the solution space in Minkowski space. We now explain this construction.
Let~$(\scrM, g)$ be the two-dimensional Minkowski space
(thus~$\scrM = \R^{1,1}$ with the metric~\eqref{mink}). Moreover, we let~$S \scrM = \scrM \times \C^2$
be the trivial spinor bundle, again with the spin scalar product~\eqref{ssprod}.
Then the inclusions
\[ \scrR \subset \scrM \qquad \text{and} \qquad S\scrR = \scrR \times \C^2
\subset \scrM \times \C^2 = S\scrM \]
are clearly isometries.
The Dirac operator and the Dirac equation are again given by~\eqref{Dirop}
and~\eqref{Direq}. For clarity, we denote the scalar product~\eqref{print}
in Minkowski space with an additional subscript~$\scrM$. For convenience, we always
choose~$\scrN$ as the Cauchy surface~$\{t=0\}$, so that
\beq \label{printM}
(\Psi | \Phi)_\scrM = 2 \pi \int_{-\infty}^\infty \Sl \Psi | \gamma^0 \Phi \Sr|_{(0,x)}\: dx
\eeq
(to avoid confusion, we consistently denote wave functions in Minkowski space by capital Greek letters,
whereas wave functions in Rindler space-time are denoted by small Greek letters).
The corresponding Hilbert space is denoted by~$(\H_\scrM, (.|.)_\scrM)$.
In order to extend Dirac solutions from Rindler space-time to Minkowski space, let~$\psi \in \Cisc(\scrR, S\scrR)$
be a solution with spatially compact support. Thus restricting it to the ray~$\{ (0,x) \text{ with } x>0\}$
gives a smooth function with compact support. We extend this function by zero to
the Cauchy surface~$\{t=0\}$, i.e.
\[ \Psi_0(x) := \left\{ \begin{array}{cl} \psi(0,x) & \text{if~$x > 0$} \\
0 & \text{if~$x \leq 0$} \end{array} \right. \quad \in C^\infty_0(\R, \C^2) \:. \] 
Solving the Cauchy problem in~$\scrM$ with initial data~$\Psi_0$ yields a solution~$\Psi(t,x)$ in Minkowski space.
We thus obtain an isometric embedding
\[ \iota_\scrM \::\: \H \rightarrow \H_\scrM \:. \]
It is also useful to introduce the operator~$\pi_\scrR$ as the restriction to Rindler space-time,
\[ \pi_\scrR \::\: \H_\scrM \rightarrow \H \:,\qquad \pi_\scrR \Psi = \Psi |_\scrR \:. \]
Obviously, the identity
\[ \pi_\scrR \circ \iota_\scrM = \1_\H \]
holds. Moreover, for every~$\Psi \in \H_\scrM$ and~$\phi \in \H$,
\[ \big( \Psi \,\big|\, \iota_\scrM \phi \big)_\scrM = 2\pi \int_0^\infty 
\Sl \Psi | \gamma^0 \phi \Sr|_{(0,x)}\: dx = \big( \pi_\scrR \Psi \,\big|\, \phi \big) \:, \]
which can be written as
\[ \iota_\scrM^* = \pi_\scrR \:. \]
This relation also shows that the orthogonal complement of the image of~$\iota_\scrM$ coincides
with the kernel of~$\pi_\scrR$, consisting of all Dirac solutions in Minkowski space
which vanish on the ray~$\{ (0,x) \text{ with } x>0\}$.

In analogy to~\eqref{stip}, the space-time inner product in Minkowski space is defined by
\[ \bra \Psi|\Phi \ket_\scrM := \int_\scrM \Sl \Psi | \Phi \Sr|_q\: d\mu_\scrM(q)\:. \]
It is not directly related to~\eqref{stip} because one integrates over a different space-time region.
However, a direct connection can be obtained by inserting the characteristic function
of Rindler space-time into the integrand,
\beq \label{stipMR}
\bra \Psi|\Phi \ket_\scrR := \int_\scrM \chi_\scrR(q) \:\Sl \Psi | \Phi \Sr|_q\: d\mu_\scrM(q)\:.
\eeq
Then for any~$\Psi, \Phi \in C^\infty_0(\scrM, S\scrM)$,
\[ \bra \Psi|\Phi \ket_\scrR = \bra \pi_\scrR \Psi \,|\, \pi_\scrR \Phi \ket \:. \]
Introducing the {\em{relative fermionic signature operator}}~$\Sig_\scrR : \D(\Sig_\scrR) \subset \H_\scrM
\rightarrow \H_\scrM$ in analogy to~\eqref{Sdef} by
\beq \label{SigRdef}
\bra \Psi | \Phi \ket_\scrR = (\Psi \,|\, \Sig_\scrR \Phi)_\scrM \qquad  \text{for all~$\Psi \in \H_\scrM$} \:,
\eeq
this fermionic signature operator in Rindler space-time is recovered by
\beq \label{Sigintrinsic}
\Sig = \pi_\scrR\, \Sig_\scrR\, \iota_\scrM \qquad \text{with} \qquad \D(\Sig) = \pi_\scrR
\big( \D(\Sig_\scrR) \big) \:.
\eeq
With this in mind, in the remainder of the paper we work exclusively in Minkowski space.
For notational simplicity, the subscript~$\scrM$ will be omitted in what follows.

\section{The Relative Fermionic Signature Operator as an Unbounded Operator} \label{secunbound}
From now on, all the objects without subscript~$\scrR$ refer to Minkowski space~$\scrM$.
\begin{Lemma} \label{lemmabound}
For every~$\Phi \in \Cisc(\scrM, S\scrM) \cap \H$, there is a constant~$c=c(\Phi)$
such that
\[ \big| \bra \Psi|\Phi \ket_\scrR \big| \leq c(\Phi)\: \|\Psi\| \qquad \text{for all~$\Psi \in \H$}\:. \]
\end{Lemma}
\Proof Let~$\Phi \in \Cisc(\scrM, S\scrM) \cap \H$. Then its restriction to the Cauchy surface~$\{t=0\}$ is compact, i.e.
\[ \supp \Phi(0,.) \subset (-R, R) \:. \]
Due to finite propagation speed, we know that
\beq \label{supppsi}
\supp \Phi(t,.) \subset (-R-|t|, R+|t|) \qquad \text{for all~$t \in \R$}\:.
\eeq
We now make use of the fact that solutions of the massive Dirac equation for compactly supported initial data
decay rapidly in null directions. More precisely, for any~$p \in \N$ there is a constant~$C=C(\Phi,p)$ such that
\beq \label{psidecay}
\big\| \Phi(t,x) \big\|_{\C^4} \leq \frac{C}{1+|t|^p} \qquad \text{for all~$t \in \R$ and~$x \geq |t|$}\:.
\eeq
This inequality can be verified in two ways. One method is to specialize the more general results
in asymptotically flat space-times as derived in~\cite{treude}.
Another method is to use that each component of~$\Phi$ is a solution of the Klein-Gordon equation
\[ \big( \partial_t^2 - \partial_x^2 + m^2 \big) \Phi(t,x) = 0 \]
and to apply the estimates in~\cite[Theorem~7.2.1]{hormanderhyp}, choosing the parameter~$N$
in this theorem to be negative and large.

Combining~\eqref{supppsi} and~\eqref{psidecay} with the Schwarz inequality, we obtain the estimate
\begin{align*}
\int_\scrR \big| \Sl \Psi | \Phi \Sr \big| \:dt\, dx 
&\leq \int_{-\infty}^\infty dt \int_{|t|}^{|t|+R} dx \; \|\Psi(t,x)\|_{\C^4}\: \|\Phi(t,x)\|_{\C^4} \\
&\leq \int_{-\infty}^\infty \; \|\Psi(t,.)\|_{L^2(dx)}\: \frac{C \sqrt{R}}{1+|t|^p}\: dt
= C\: \sqrt{R} \: \|\Psi\| \int_{-\infty}^\infty \frac{dt}{1+|t|^p} \:.
\end{align*}
Choosing~$p=2$ gives the desired estimate.
\QED

Using this lemma, for any~$\Psi \in \Cisc(\scrM, S\scrM) \cap \H$, the Fr{\'e}chet-Riesz theorem
gives a unique vector~$\Sig_\scrR \Phi \in \H$ such that~\eqref{SigRdef} holds. This makes it possible
to introduce the relative fermionic signature operator as the densely defined operator
\beq \label{SigRdense}
\Sig_\scrR \::\: \Cisc(\scrM, S\scrM) \cap \H \rightarrow \H \:.
\eeq
From~\eqref{SigRdef} it is obvious that~$\Sig_\scrR$ is symmetric, i.e.
\[ ( \Psi \,|\, \Sig_\scrR \Phi ) = (\Sig_\scrR \Psi \,|\, \Phi ) \qquad
\text{for all~$\Psi, \Phi \in \Cisc(\scrM, S\scrM) \cap \H$} \:. \]

We point out that the operator~$\Sig_\scrR$ is {\em{unbounded}}. This can be understood from the
fact that the inequality~\eqref{psidecay} and the subsequent estimate depend essentially on the
support of~$\Phi$. In particular, if we consider a sequence of wave functions~$(\Phi_n)_{n \in \N}$ whose
support is shifted more and more to the right,
\beq \label{Phin}
\Phi_n(t,x) = \Phi(t, x-n) \:,
\eeq
then the constant~$c(\Phi_n)$ in the statement of Lemma~\ref{lemmabound} must be chosen larger
and larger if~$n$ is increased. This indicates that the inequality
\beq \label{violate}
\big| \bra \Psi|\Phi \ket_\scrR \big| \leq c\:\|\Phi\|\, \|\Psi\| \quad \text{for all~$\Psi,\Phi \in \H$} 
\qquad \text{is violated}\:,
\eeq
no matter how large the constant~$c$ is chosen
(for a formal proof of this statement see Corollary~\ref{corminfinite} below). 
%
Using the terminology introduced in~\cite[Section~3.2]{finite}, 
the statement~\eqref{violate} means that {\em{Rindler space-time is not $m$-finite}}.
Nevertheless, the estimate of Lemma~\ref{lemmabound} enables us to introduce
the fermionic signature operator as a densely defined, unbounded symmetric operator.
This makes it unnecessary to employ the mass oscillation methods introduced in~\cite{infinite} 
for the construction of the fermionic signature operator in space-times of infinite lifetime.

In order to get into the position to use spectral methods, we must construct a self-adjoint extension of
the relative fermionic signature operator. Our method is to compute~$\Sig_\scrR$ in more detail in
momentum space. As we shall see, working with plane waves in a suitable parametrization
in momentum space, the operator~$\Sig_\scrR$
becomes a multiplication operator, making it possible to construct a self-adjoint extension
with standard functional analytic methods.

\section{Transformation to Momentum Space}
For the following computations, it is most convenient to work in momentum space.
We denote the position and momentum variables by~$q=(t,x)$ and~$p = (\omega, k)$, respectively.
Clearly, any smooth and spatially compact Dirac solution~$\Psi \in \Cisc(\scrM, S\scrM)$ can be represented as
\beq \label{fourier0}
\Psi(q) = \int_{\R^2} \frac{d^2p}{(2 \pi)^2}\:\hat{\Psi}(p)\: \delta(p^2-m^2)\: e^{-ipq} \:,
\eeq
where~$\hat{\Psi}$ is a smooth function on the mass shell (and~$pq = \omega t - kx$ is the Minkowski inner product).
In this momentum representation,
the Dirac equation~\eqref{Direq} reduces to the algebraic equation
\[ \big( \slashed{p} - m \big) \hat{\Psi}(p) = 0 \:. \]
The matrix~$\slashed{p} - m$ has eigenvalues~$0$ and~$-2m$. Its kernel
is positive definite with respect to the spin scalar product if~$p$ is on the upper mass shell,
and it is negative definite if~$p$ is on the lower mass shell.
Thus we can choose a spinor~$\f(p)$ with the properties
\beq \label{fdef}
(\slashed{p}-m) \,\f(p) = 0 \qquad \text{and} \qquad
\Sl \f(p) | \f(p) \Sr = \epsilon(\omega) \:,
\eeq
where~$\epsilon$ is the sign function~$\epsilon(\omega)=1$
for $\omega \geq 0$ and $\epsilon(\omega)=-1$ otherwise. More specifically, we choose
\beq \label{fdefex}
\f(p) = \frac{1}{\sqrt{2m}} \, \frac{1}{\sqrt{\epsilon(\omega)\, (\omega-k)}}\:
\begin{pmatrix} m \\ \omega - k \end{pmatrix} \:.
\eeq
\begin{Lemma} \label{lemmaspinor}
The spinor~$\f(p)$ satisfies the relations
\begin{align*}
\fiberpairing{ \f(\omega,k)}{ \gamma^0 \,{\f}(-\omega,k) } &=0 \\
\fiberpairing{ \f(\omega,k)}{ \gamma^0 \,{\f}(\omega,k) } &= \frac{|\omega|}{m}\:.
\end{align*}
\end{Lemma}
\Proof These relations can be verified in a straightforward manner using the explicit formulas~\eqref{fdef}
and~\eqref{gammarep}. Alternatively, they can also be derived abstractly by applying
the anti-commutation relations of the Dirac matrices:
\begin{align*}
\fiberpairing{ &\f(\omega,k)}{ \gamma^0 \,{\f}(-\omega,k) } =
\frac{1}{m} \:\fiberpairing{ \slashed{p}\, \f(\omega,k)}{ \gamma^0 \,{\f}(-\omega,k) } \\
&= \frac{1}{m} \:\fiberpairing{ \f(\omega,k)}{ \big(\omega \gamma^0 - k \gamma^1 \big)\, \gamma^0 \,{\f}(-\omega,k) } \\
&= \frac{1}{m} \:\fiberpairing{ \f(\omega,k)}{ \gamma^0 \,\big(\omega \gamma^0 + k \gamma^1 \big) \,{\f}(-\omega,k) }
= -\fiberpairing{ \f(\omega,k)}{ \gamma^0 \,{\f}(-\omega,k) } \\
\fiberpairing{ &\f(\omega,k)}{ \gamma^0 \,{\f}(\omega,k) } =
\frac{1}{m} \:\fiberpairing{ \slashed{p}\, \f(\omega,k)}{ \gamma^0 \,{\f}(\omega,k) } \\
&= \frac{1}{m} \:\fiberpairing{ \f(\omega,k)}{ \gamma^0 \,\big(\omega \gamma^0 + k \gamma^1 \big) {\f}(\omega,k) } \\
&= \frac{2 \omega}{m} \:\fiberpairing{ \f(\omega,k)}{  {\f}(\omega,k) }
- \frac{1}{m} \:\fiberpairing{ \f(\omega,k)}{ \gamma^0 \,\slashed{p} {\f}(\omega,k) } \\
&= \frac{2 \omega}{m} \:\fiberpairing{ \f(\omega,k)}{  {\f}(\omega,k) }
- \fiberpairing{ \f(\omega,k)}{ \gamma^0 \,{\f}(\omega,k) } \:.
\end{align*}
Using the right relation in~\eqref{fdef}, the result follows.
\QED

It is convenient to represent the spinor~$\hat{\Psi}(p)$ in~\eqref{fourier0} as a complex multiple of the spinor~$\f(p)$.
Thus we write the Fourier integral~\eqref{fourier0} as
\beq \label{fourierrep}
\Psi(q) = \int_{\R^2} \frac{d^2p}{2 \pi}\:\epsilon(\omega) \: \delta(p^2-m^2)\: g(p)  \:\f(p)\: e^{-ipq}
\eeq
with a complex-valued function~$g(p)$.
In the next two lemmas we specify the regularity of the function~$g(p)$ and
rewrite the scalar product~\eqref{printM} in momentum space.

\begin{Lemma} \label{lemmaschwartz}
For every smooth and spatially compact Dirac solution~$\Psi \in \Cisc(\scrM, S\scrM)$,
the function~$g$ in the representation~\eqref{fourierrep} is a Schwartz function on the mass shells, i.e.
\[ g_\pm(k) := g \big(\pm \sqrt{k^2+m^2}, k\big) \;\in\; {\mathcal{S}}(\R, \C) \:. \]
\end{Lemma}
\Proof Evaluating~\eqref{fourierrep} at~$q^0=0$ gives
\begin{align*}
\Psi(0,x) &= \int_{\R^2} \frac{d^2p}{2 \pi}\:\epsilon(\omega) \: \delta(p^2-m^2)\: g(p)  \:\f(p)\: e^{ikx} \notag \\
&= \frac{1}{2}
\int_{-\infty}^\infty \frac{dk}{2 \pi}\:\sum_\pm \frac{\epsilon(\omega)}{\sqrt{k^2+m^2}}\: g(p)  \:\f(p)\: e^{ikx}
\Big|_{p=\big(\pm \sqrt{k^2+m^2}, k \big)} \\
i \partial_t\Psi(0,x) &= \int_{\R^2} \frac{d^2p}{2 \pi}\:\epsilon(\omega) \: p_0\: \delta(p^2-m^2)\: g(p)  \:\f(p)\: e^{ikx} 
\notag \\
&= \frac{1}{2} \int_{-\infty}^\infty \frac{dk}{2 \pi}\:\sum_\pm g(p)  \:\f(p)\: e^{ikx}
\Big|_{p=\big(\pm \sqrt{k^2+m^2}, k \big)} \:.
\end{align*}
On the other hand, taking the one-dimensional Fourier transform, we know that
\[ \Psi(0,x) = \int_{-\infty}^\infty \frac{dk}{2 \pi}\:\hat{\Phi}_0(k)\: e^{ik x} \qquad \text{and} \qquad
\partial_t \Psi(0,x) = \int_{-\infty}^\infty \frac{dk}{2 \pi} \:\hat{\Phi}_1(k)\: e^{ik x} \]
for Schwartz functions~$\hat{\Phi}_0, \hat{\Phi}_1 \in {\mathcal{S}}(\R, \C^2)$. Comparing the integrands, we obtain
\beq \label{gpmdef}
\pm \sqrt{k^2+m^2}\: \hat{\Phi}_0(k) + \hat{\Phi}_1(k) = g_\pm(k)  \:\f\big(\pm \sqrt{k^2+m^2}, k \big) \:.
\eeq
Taking the spin scalar product with~$\f$ and using the right equation in~\eqref{fdef}, we get
\[ g_\pm(k) = \Sl \Big(
\sqrt{k^2+m^2}\: \hat{\Phi}_0(k) \pm \hat{\Phi}_1(k) \Big) \:\big|\: \f\big(\pm \sqrt{k^2+m^2}, k \big) \Sr \:. \]
According to~\eqref{fdefex}, the spinor~$\f$ is smooth and grows at most linearly for large~$k$
(meaning that~$\|\f\|_{\C^2} \leq c(1+|k|)$ for a suitable constant~$c$).
This gives the result.
\QED

\begin{Lemma} In the Fourier representation~\eqref{fourierrep}, the
scalar product~\eqref{printM} can be written as
\beq
\scalarproduct{\Psi}{\tilde{\Psi}} = \frac{1}{2m}
\int_{\R^2} \overline{g(p)} \:\tilde{g}(p)\: \delta \big( p^2-m^2 \big)\: d^2p
\:\label{printmom} \:.
\eeq
\end{Lemma}
\Proof We substitute~\eqref{fourierrep} into~\eqref{print}. 
In view of the rapid decay of~$g$ (see Lemma~\ref{lemmaschwartz}), we may commute the
integrals using Plancherel's theorem to obtain
\begin{align*}
\scalarproduct{\Psi}{\tilde{\Psi}} &= 2 \pi \int_{-\infty}^\infty dx\,
\int_{\R^2} \frac{d^2p}{2 \pi}\: \epsilon(\omega) \:\delta \big( p^2-m^2 \big)\: \overline{g(p)}  \\
&\qquad \times \int_{\R^2} \frac{d^2\tilde{p}}{2 \pi}\:\epsilon(\tilde{\omega}) \:
\delta \big( \tilde{p}^{\,2}-m^2 \big)\: \tilde{g}(\tilde{p}) \:
\fiberpairing{ \f(p)}{ \gamma^0 \,{\f}(\tilde{p})} \:e^{-i (k-\tilde{k}) x} \\
&= 2 \pi \int_{\R^2} \frac{d^2p}{2 \pi}\: \epsilon(\omega) \:\delta \big( p^2-m^2 \big)\: \overline{g(p)} 
\int_{\R^2} \frac{d^2\tilde{p}}{2 \pi}\:\epsilon(\tilde{\omega}) \:\delta \big( \tilde{p}^{\,2}-m^2 \big) \\
&\qquad \times 2 \pi \delta \big( k - \tilde{k} \big)\: \tilde{g}(\tilde{p}) \:
\fiberpairing{ \f(p)}{ \gamma^0 \,{\f}(\tilde{p})} \\
&= \int_{\R^2} d^2p\: \epsilon(\omega) \:\delta \big( p^2-m^2 \big)\: \overline{g(p)} 
\int_{\R^2} d\tilde{\omega}\:\epsilon(\tilde{\omega}) \:\delta \big( \tilde{\omega}^2 - k^2 -m^2 \big) \\
&\qquad \times \tilde{g}\big( \tilde{\omega}, k \big) \:
\fiberpairing{ \f(p)}{ \gamma^0 \,{\f}\big( \tilde{\omega}, k \big)} \\
&= \int_{\R^2} d^2p\: \epsilon(\omega) \:\delta \big( p^2-m^2 \big)\: \overline{g(p)} \;
\frac{1}{2|\omega|} \\
&\qquad \times \sum_\pm \epsilon(\pm\omega)\: \tilde{g}\big( \pm \omega, k \big) \:
\fiberpairing{ \f(p)}{ \gamma^0 \,{\f}\big( \pm \omega, k \big)}\:.
\end{align*}
Applying Lemma~\ref{lemmaspinor} gives~\eqref{printmom}.
\QED

We finally choose a convenient parametrization of the mass shells:
\begin{Prp} \label{prpsprod}
In the parametrization
\beq \label{parameter}
\begin{pmatrix} \omega \\ k \end{pmatrix} = ms \begin{pmatrix} \, \cosh \alpha \\ \sinh \alpha \end{pmatrix} 
\qquad \text{with} \qquad s \in \{\pm 1\} \text{ and }\, \alpha \in \R\:,
\eeq 
the scalar product~\eqref{printM} takes the form
\beq
\scalarproduct{\Psi}{\tilde{\Psi}} = \frac{1}{4 m}
\sum_{s =\pm 1} \int_{-\infty}^\infty  \overline{g(s,\alpha)}\: \tilde{g}(s,\alpha) \:d\alpha
\:\label{scalar product1} \:.
\eeq
\end{Prp} \noindent
We remark that the variable~$\alpha$ has the interpretation as the rapidity
of the wave in the rest frame of the observer.
\Proof[Proof of Proposition~\ref{prpsprod}.] We carry out the $\omega$-integration in~\eqref{printmom},
\begin{align*}
\int_{\R^2} & \overline{g(p)} \:\tilde{g}(p)\:\delta \big( p^2-m^2 \big)\: d^2p
= \sum_{\pm} \int_{-\infty}^\infty \frac{dk}{2 \sqrt{k^2+m^2}} \:\big( \overline{g} \:\tilde{g} \big)\big|_{\big(\pm \sqrt{k^2+m^2}, k \big)} \\
&= \sum_{s=\pm}
\int_{-\infty}^\infty m\, \cosh \alpha\; \frac{1}{2 m \cosh \alpha}
\:\big( \overline{g} \:\tilde{g} \big)\big|_{\big(ms \cosh \alpha, m s \sinh \alpha \big)} \:d\alpha \\
&= \frac{1}{2} \:\sum_{s=\pm}
\int_{-\infty}^\infty\:\overline{g(s,\alpha)}\: \tilde{g}(s,\alpha)\: d\alpha\:.
\end{align*}
This gives the result.
\QED

\section{The Relative Fermionic Signature Operator in Momentum Space}
In this section, we compute the fermionic signature operator more explicitly in momentum space.
The first step is to transform the space-time inner product to momentum space.

\begin{Prp}
For any~$\Psi, \tilde{\Psi} \in \Cisc(\scrM, S\scrM) \cap \H$, the space-time inner product~\eqref{stipMR}
takes the form
\begin{align}
\bracket{\Psi}{\tilde{\Psi}}_\scrR &=  
\frac{1}{4 m} \sum_{s,\tilde{s}=\pm1} \int_{-\infty}^\infty
d \alpha \; \lim_{\varepsilon \searrow 0} \int_{-\infty}^\infty d \tilde{\alpha} \;
I_\varepsilon\big(s,\alpha; \tilde{s}, \tilde{\alpha} \big) \;
\:\overline{g(s,\alpha)}  
\;\tilde{g}(\tilde{s}, \tilde{\alpha}) \:, \label{bracket1} 
\end{align}
where~$I_\varepsilon$ is the kernel
\beq \label{Iform}
I_\varepsilon \big( s,\alpha; \tilde{s}, \tilde{\alpha} \big) = \frac{1}{4 \pi^2 \,m} \times
\left\{ \begin{array}{cl} \displaystyle \frac{s \cosh \beta}{1 - \cosh(2 \beta + i \varepsilon s)} & \text{if~$s = \tilde{s}$}  \\[1.0em]
\displaystyle -\frac{s \sinh \beta}{1 + \cosh (2 \beta)} & \text{if~$s \neq \tilde{s}$}
\end{array} \right.
\eeq
and
\beq \label{betadef}
\beta := \frac{1}{2}\: \big(\alpha-\tilde{\alpha} \big)\:.
\eeq
\end{Prp}
\begin{proof} Using the Fourier representation~\eqref{fourierrep} in~\eqref{stipMR}, we obtain
\begin{align}
\bra \Psi | \tilde{\Psi} \ket_\scrR &= \int_{\scrM} dt\, dx\:\chi_\scrR(t,x)
\int_{\R^2} \frac{d^2p}{2 \pi}\:\epsilon(\omega) \: \delta(p^2-m^2)\: \overline{g(p)} \notag \\
&\qquad \times \int_{\R^2} \frac{d^2\tilde{p}}{2 \pi}\: \epsilon(\tilde{\omega}) \: \delta(\tilde{p}^{\,2}-m^2)\: \tilde{g}(\tilde{p}) \:
\fiberpairing{ \f(p)}{ {\f}(\tilde{p}) } \:e^{i (p-\tilde{p}) q} \notag \\
&= \int_{\R^2} \frac{d^2p}{(2 \pi)^2}\: \epsilon(\omega) \:\delta(p^2-m^2)\: \overline{g(p)} \notag  \\
&\qquad \times \int_{\R^2} {d^2\tilde{p}}\:\epsilon(\tilde{\omega}) \: \delta(\tilde{p}^{\,2}-m^2)\: \tilde{g}(\tilde{p}) \:
\fiberpairing{ \f(p) }{ {\f}(\tilde{p}) } \:K(p, \tilde{p}) \:, \label{Kres}
\end{align}
where the kernel~$K(p, \tilde{p})$ is defined by
\beq \label{Kdef}
K(p, \tilde{p}) = \int_{\scrM} \chi_\scrR(t,x)\: e^{i (p-\tilde{p}) q}\: dt\,dx  \:.
\eeq
Rewriting the integrals in~\eqref{Kres} in the parametrization~\eqref{parameter}
(exactly as in the proof of Proposition~\ref{prpsprod}), we get
\beq \label{tocompute}
\bra \Psi | \tilde{\Psi} \ket_\scrR = 
\frac{1}{16 \pi^2} \sum_{s,\tilde{s}=\pm1} \int_{-\infty}^\infty d \alpha \int_{-\infty}^\infty d \tilde{\alpha} \;s\, \tilde{s}\,
\:\overline{g(s,\alpha)}  \: \tilde{g}(\tilde{s}, \tilde{\alpha})\:
\fiberpairing{ \f(p) }{ {\f}(\tilde{p}) } \:K(p, \tilde{p})\:.
\eeq
Applying Lemma~\ref{lemmasprod} and Lemma~\ref{lemmaKcompute} below, the result follows.
\QED

Comparing~\eqref{scalar product1} and~\eqref{bracket1}, one can immediately read off the
relative fermionic signature operator as defined by~\eqref{SigRdef} and~\eqref{SigRdense}:
\begin{Corollary} \label{corS}
For any~$\tilde{\Psi} \in \Cisc(\scrM, S\scrM) \cap \H$,
\[ \big(\Sig_\scrR \tilde{\Psi} \big)(s, \alpha) =
\sum_{\tilde{s}=\pm1} \lim_{\varepsilon \searrow 0} \int_{-\infty}^\infty
I_\varepsilon \big(s,\alpha; \tilde{s}, \tilde{\alpha} \big) \;\tilde{g}(\tilde{s}, \tilde{\alpha})\:  d \tilde{\alpha} \:. \]
\end{Corollary}

In the following two lemmas we compute the spin scalar product and the
kernel in~\eqref{tocompute}.
\begin{Lemma} \label{lemmasprod}
In the parametrization~\eqref{parameter},
the spin scalar product of the spinors~\eqref{fdefex} is computed by
\[ \Sl \f(s,\alpha) \,|\, \f(\tilde{s}, \tilde{\alpha}) \Sr = \left\{
\begin{array}{cl} s \cosh \beta & \text{if~$s=\tilde{s}$} \\
s \sinh \beta & \text{if~$s \neq \tilde{s}$}\:.
\end{array}  \right. \]
\end{Lemma}
\Proof Using~\eqref{fdefex} and~\eqref{ssprod}, we have
\[ \Sl \f(p) | \f(\tilde{p}) \Sr
= \frac{1}{2} \, \frac{(\omega - k) + (\tilde{\omega} - \tilde{k})}{\sqrt{\epsilon(\omega)\, (\omega-k)\: \epsilon(\tilde{\omega})\, (\tilde{\omega}-\tilde{k})}} \:. \]
In the parametrization~\eqref{parameter}, we obtain
\begin{align*}
\Sl \f(s,\alpha) | \f(\tilde{s}, \tilde{\alpha}) \Sr
&= \frac{1}{2} \, \frac{s e^{-\alpha} + \tilde{s} e^{-\tilde{\alpha}}}{e^{-\frac{\alpha}{2} - \frac{\tilde{\alpha}}{2}}}
= \frac{1}{2} \, \Big( s e^{\beta} + \tilde{s} e^{-\beta} \Big)\:.
\end{align*}
This gives the result.
\QED

\begin{Lemma} \label{lemmaKcompute}
In the parametrization~\eqref{parameter}, the distribution~$K(p, \tilde{p})$ defined by~\eqref{Kdef} has the form
\[ K(s,\alpha; \tilde{s}, \tilde{\alpha}) = \frac{1}{m^2} \times
\left\{ \begin{array}{cl} \displaystyle \lim_{\varepsilon \searrow 0} \frac{1}{1 - \cosh(\alpha-\tilde{\alpha} - i \varepsilon s)} & \text{if~$s = \tilde{s}$}  \\[1.0em]
\displaystyle \frac{1}{1 + \cosh (\alpha-\tilde{\alpha})} & \text{if~$s \neq \tilde{s}\:.$}
\end{array} \right. \]
\end{Lemma}
\Proof We first write~\eqref{Kdef} as
\beq
K(p, \tilde{p}) = \int_{\R^{2}} dt\,dx \: \chi(x - t) \:\chi( x + t) \: e^{i (p-\tilde{p}) q} \:.  \label{eq:K}
\eeq
Introducing null coordinates
\[ u = \frac{1}{2}\:( t-x ) \qquad \text{and} \qquad v = \frac{1}{2} \:( t+x ) \]
as well as corresponding momenta
\[ p_u =\omega -\tilde{\omega} + k - \tilde{k} \qquad \text{and} \qquad
p_v =\omega-\tilde{\omega} - k + \tilde{k} \:, \]
we can compute the integrals in~\eqref{eq:K} to obtain
\begin{align*}
 K(p_u,p_v ) &= 2 \int_{\R^{2}} du\, dv  \, \chi(-2u) \:\chi(2v) \: e^{i ( p_u u + p_v v)} =
2 \int_{-\infty}^0 du \: e^{i p_u u}\: \int_0^\infty dv \: e^{i  p_v v} \\
&=2 \lim_{\varepsilon \searrow 0} \int_{-\infty}^0 du \: e^{i p_u u + \varepsilon u}\: 
\lim_{\varepsilon' \searrow 0}  \int_0^\infty dv \: e^{i  p_v v - \varepsilon' v} 
=  2 \lim_{\varepsilon, \varepsilon' \searrow 0 } \frac{1}{p_u- i \varepsilon} \: \frac{1}{p_v + i \varepsilon'} \:.
\end{align*}

We next express~$p_u$ in the parametrization~\eqref{parameter},
\begin{align*}
p_u &= (\omega+k) - (\tilde{\omega} + \tilde{k})
= m s \,\big(\cosh (\alpha) + \sinh (\alpha)) - m \tilde{s} (\cosh (\tilde{\alpha}) + \sinh (\tilde{\alpha}) \big) \\
&= m \big( s e^\alpha - \tilde{s} e^{\tilde{\alpha}} \big) \:.
\end{align*}
This gives
\begin{equation}\label{function}
\lim_{\varepsilon \searrow 0} \frac{1}{p_u- i \varepsilon}
=  \lim_{\varepsilon \searrow 0} \frac{1}{m ( s e^\alpha - \tilde{s} e^{\tilde{\alpha}} ) - i \varepsilon} \:.
\end{equation}
We distinguish the two cases~$s \neq \tilde{s}$ and~$s = \tilde{s}$. In the case~$s \neq \tilde{s}$,
the denominator in~\eqref{function} is always non-zero.
Therefore, we can take the limit~$\varepsilon \searrow 0$ pointwise to obtain
\begin{align*}
\lim_{\varepsilon \searrow 0} \frac{1}{p_u- i \varepsilon} = \frac{1}{m s}
\:\frac{1}{e^\alpha + e^{\tilde{\alpha}}} = \frac{e^{-\alpha}}{m s} \:\frac{1}{1 + e^{-2 \beta}}
\qquad (s \neq \tilde{s})\:,
\end{align*}
where~$\beta$ is again given by~\eqref{betadef}.
In the remaining case~$s = \tilde{s}$,
we rewrite~\eqref{function} as
\begin{align*}
\lim_{\varepsilon \searrow 0} \frac{1}{p_u- i \varepsilon} &= \frac{1}{ms} \:
\lim_{\varepsilon \searrow 0} \frac{1}{(e^\alpha - e^{\tilde{\alpha}} ) - i \varepsilon s/m}
=  \frac{e^{-\alpha}}{ms}  \:\lim_{\varepsilon \searrow 0} \frac{1}{1 - e^{-2 \beta} - i \varepsilon s e^{-\alpha}/m} \\
&
= \frac{e^{-\alpha}}{ms}  \:\lim_{\delta \searrow 0} \frac{1}{1 - e^{-2 \beta} - i \delta s\, e^{-2 \beta}} 
= \frac{e^{-\alpha}}{ms}  \:\lim_{\delta \searrow 0} \frac{1}{1 - e^{-2 \beta+i \delta s}} 
\:,
\end{align*}
where~$\delta = \varepsilon e^{-\alpha+2 \beta}/m > 0$.
We conclude that
\begin{align*}
\lim_{\varepsilon \searrow 0} \frac{1}{p_u- i \varepsilon} =
\frac{e^{-\alpha}}{ms}  \lim_{\delta \searrow 0} \frac{1}{1 - e^{-2 \beta+i \delta s}} 
\qquad (s = \tilde{s})\ .
\end{align*}

Treating~$p_v$ in the same way, we obtain
\begin{align}
\lim_{\varepsilon \searrow 0} \frac{1}{p_u - i \varepsilon}
&= \left\{ \begin{array}{cl} \displaystyle \frac{e^{-\alpha}}{m s} \:\frac{1}{1 + e^{-2 \beta}} & \text{if~$s \neq \tilde{s}$}
\\[0.8em]
\displaystyle \frac{e^{-\alpha}}{ms}  \lim_{\varepsilon \searrow 0} \frac{1}{1 - e^{-2 \beta+i \varepsilon s}} & \text{if~$s = \tilde{s}$} \end{array} \right. \label{1}\\
\lim_{\varepsilon' \searrow 0} \frac{1}{p_v + i \varepsilon'}
&= \left\{ \begin{array}{cl} \displaystyle \frac{e^{\alpha}}{m s} \:\frac{1}{1 + e^{2 \beta}} & \text{if~$s \neq \tilde{s}$}
\\[0.8em]
\displaystyle \frac{e^{\alpha}}{ms}  \lim_{\varepsilon' \searrow 0} \frac{1}{1 - e^{2 \beta-i \varepsilon' s}} & \text{if~$s = \tilde{s}\:.$} \end{array} \right.\label{2}
\end{align}
When multiplying~\eqref{1} and~\eqref{2}, the fact that both limits~$\varepsilon, \varepsilon' \searrow 0$
exist in the distributional sense justifies that we can simply set~$\varepsilon= \varepsilon'$ and take the limit.
Using~\eqref{betadef}, the result follows.
\QED

\section{Diagonalizing the Relative Fermionic Signature Operator}
In Corollary~\ref{corS}, the relative fermionic signature operator~$\Sig_\scrR$
was represented by an integral operator. Since the kernel~$I(s,\alpha; \tilde{s}, \tilde{\alpha})$
only depends on the difference~$\alpha-\tilde{\alpha}$ (see~\eqref{Iform} and~\eqref{betadef}), 
we can diagonalize the fermionic operator with the plane wave ansatz
\beq \label{planewave}
g \big( \tilde{s}, \tilde{\alpha} \big) = e^{-i \ell \tilde{\alpha}} \: \hat{g}(\tilde{s}, \ell)
\eeq
for a real parameter~$\ell$
(thus~$\ell$ is the variable conjugate to the rapidity; to our knowledge it does not have a
straightforward physical interpretation).
Clearly, the plane wave is not a vector in our Hilbert space~$\H$. But the corresponding spectral parameter
corresponds to a point in the continuous spectrum of~$\Sig_\scrR$.
For clarity, we first give the computations. The functional analytic framework will
be developed in Section~\ref{secFA} below.
\begin{Lemma} \label{lemmadiagonal} The integral kernel~$I_\varepsilon$, \eqref{Iform}, satisfies the relation
\[ \lim_{\varepsilon \searrow 0} \int_{-\infty}^\infty
I_\varepsilon \big(s,\alpha; \tilde{s}, \tilde{\alpha} \big)\: e^{-i \ell \tilde{\alpha}}\: d\tilde{\alpha}
= e^{-i \ell \alpha} \: \frac{\ell}{2 \pi m} \times
\left\{ \begin{array}{cl} \displaystyle \frac{2}{1 + e^{-2\pi s \ell}} & \text{if~$s = \tilde{s}$}  \\[1.0em]
\displaystyle -\frac{i s}{\cosh(\pi \ell)} & \text{if~$s \neq \tilde{s}\:.$}
\end{array} \right. \]
\end{Lemma}

Using the result of this lemma, one sees that for the plane wave ansatz~\eqref{planewave},
the equation~$\Sig_\scrR g = \lambda g$ reduces to the eigenvalue equation
for a Hermitian matrix,
\beq \label{hatSdef}
\hat{\Sig}_\scrR(\ell) \,\hat{g}(\ell) = \lambda \hat{g}(\ell)
\qquad \text{with} \qquad \hat{\Sig}_\scrR(\ell) = \frac{\ell}{\pi m}
\begin{pmatrix} \displaystyle \frac{1}{1 + e^{-2\pi \ell}} & \displaystyle -\frac{i}{2 \cosh(\pi \ell)} \\[.9em]
\displaystyle \frac{i}{2 \cosh(\pi \ell)} & \displaystyle \frac{1}{1 + e^{2\pi \ell}}
\end{pmatrix} \:,
\eeq
where~$\hat{g}(\ell) \in \C^2$ is the vector with components~$\hat{g}(1, \ell)$ and~$\hat{g}(-1, \ell)$.
The matrix~$\hat{\Sig}_\scrR(\ell)$ has the eigenvalues
\beq \label{eigenvals}
\lambda=0 \qquad \text{and} \qquad \lambda = \frac{\ell}{\pi m}
\eeq
with respective eigenfunctions
\beq \label{eigenvectors}
\hat{g}(\ell) = \begin{pmatrix} i e^{-\pi \ell} \\ 1 \end{pmatrix} \qquad \text{and} \qquad
\hat{g}(\ell) = \begin{pmatrix} -i e^{\pi \ell} \\ 1 \end{pmatrix} \:.
\eeq

\Proof[Proof of Lemma~\ref{lemmadiagonal}] In the case~$s\neq \tilde{s}$, we have
\begin{align*}
\lim_{\varepsilon \searrow 0} & \int_{-\infty}^\infty
I_\varepsilon \big(s,\alpha; \tilde{s}, \tilde{\alpha} \big)\: e^{-i \ell (\tilde{\alpha}-\alpha)}\: d\tilde{\alpha} \\
&= 2 \lim_{\varepsilon \searrow 0} \int_{-\infty}^\infty
I_\varepsilon \big(s,\alpha; \tilde{s}, \tilde{\alpha} \big)\: e^{2 i \ell \beta}\: d\beta
= -\frac{s}{2\pi^2 \,m} \int_{-\infty}^\infty
\frac{\sinh \beta}{1 + \cosh (2 \beta)} \: e^{2 i \ell \beta}\: d\beta \:.
\end{align*}
The integral can be computed as follows. First, using the transformation
\[ \frac{\sinh \beta}{1 + \cosh (2 \beta)} = \frac{e^\beta - e^{-\beta}}{(e^{\beta} + e^{-\beta})^2} 
= -\frac{d}{d\beta} \bigg( \frac{1}{e^{\beta} + e^{-\beta}} \bigg) \:, \]
we can integrate by parts to obtain
\[ \int_{-\infty}^\infty
\frac{\sinh \beta}{1 + \cosh (2 \beta)} \: e^{2 i \ell \beta}\: d\beta
= 2 i \ell \int_{-\infty}^\infty
\frac{e^{2 i \ell \beta}}{e^{\beta} + e^{-\beta}} \: d\beta
= i \ell \int_{-\infty}^\infty
\frac{e^{2 i \ell \beta}}{\cosh \beta} \: d\beta
\:. \]
Now the integral can be calculated with residues. The variable transformation~$\beta \mapsto -\beta$
shows that the last integral is even in~$\ell$. Therefore, it suffices to consider the case~$\ell>0$.
Then we can close the contour in the upper half plane.
There the integrand has poles at~$\beta_n= i \pi (n+\frac{1}{2})$ with~$n \in \N_0$. This gives
\begin{align}
&\int_{-\infty}^\infty
\frac{\sinh \beta}{1 + \cosh (2 \beta)} \: e^{2 i \ell \beta}\: d\beta
= -2 \pi \ell \sum_{n=0}^\infty \text{Res} \Big( \frac{e^{2 i \ell \beta}}{\cosh \beta}, \beta_n \Big) \notag \\
&\;\;= -2 \pi \ell \sum_{n=0}^\infty (-i) \,(-1)^n \, e^{-2 \pi \ell\, (n+\frac{1}{2})}
= 2 \pi i \ell \,e^{-\pi \ell} \sum_{n=0}^\infty \big( -e^{-2 \pi \ell} \big)^n \notag \\
&\;\;=  2 \pi i \ell \,e^{-\pi \ell} \; \frac{1}{1+e^{-2 \pi \ell}} 
= \frac{i \pi \ell}{\cosh (\pi \ell)} \:. \label{int1}
\end{align}

In the case~$s=\tilde{s}$, we find similarly
\begin{align*}
\lim_{\varepsilon \searrow 0} & \int_{-\infty}^\infty
I_\varepsilon \big(s,\alpha; \tilde{s}, \tilde{\alpha} \big)\: e^{-i \ell (\tilde{\alpha}-\alpha)}\: d\tilde{\alpha}
= \frac{s}{2 \pi^2 \,m} \lim_{\varepsilon \searrow 0} \int_{-\infty}^\infty
\frac{\cosh (\beta- \frac{i \varepsilon s}{2})}{1 - \cosh(2 \beta - i \varepsilon s)}\: e^{2 i \ell \beta}\: d\beta\:.
\end{align*}
Rewriting the integrand as
\[ \frac{\cosh (\beta-\frac{i \varepsilon s}{2})}{1 - \cosh(2 \beta - i \varepsilon s)}
= -\frac{e^{\beta- \frac{i \varepsilon s}{2}} + e^{-\beta+\frac{i \varepsilon s}{2}}}{(e^{\beta-\frac{i \varepsilon s}{2}}
- e^{-\beta+\frac{i \varepsilon s}{2}})^2} = \frac{d}{d\beta} \bigg( \frac{1}{e^{\beta-\frac{i \varepsilon s}{2}}
- e^{-\beta+\frac{i \varepsilon s}{2}}} \bigg)
\:, \]
we can again integrate by parts to obtain
\beq
\int_{-\infty}^\infty
\frac{\cosh (\beta-\frac{i \varepsilon s}{2})}{1 - \cosh(2 \beta - i \varepsilon s)}\: e^{2 i \ell \beta}\: d\beta
= -i \ell \int_{-\infty}^\infty
\frac{e^{2 i \ell \beta}}{\sinh(\beta-\frac{i \varepsilon s}{2})}\: d\beta \:. \label{int2}
\eeq
Now the last integral is odd under the joint transformations
\[ \ell \mapsto -\ell \qquad \text{and} \qquad s \mapsto -s \:. \]
Therefore, it again suffices to consider the case~$\ell>0$, where the contour
can be closed in the upper half plane. In the case~$s=1$, the contour encloses the poles
at the points~$\beta_n= i \pi n$ with~$n \in \N_0$. This gives
\begin{align*}
&\lim_{\varepsilon \searrow 0} \int_{-\infty}^\infty
\frac{\cosh \beta}{1 - \cosh(2 \beta - i \varepsilon)}\: e^{2 i \ell \beta}\: d\beta
= 2 \pi \ell \sum_{n=0}^\infty \text{Res} \Big( \frac{e^{2 i \ell \beta}}{\sinh \beta} ,
\beta_n \Big) \\
&\;\;= 2 \pi \ell \sum_{n=0}^\infty \big( - e^{-2 \pi \ell} \big)^n
= \frac{2 \pi \ell}{1 + e^{-2\pi\ell}} .
\end{align*}
In the case~$s=-1$, the contour does not enclose the pole at~$\beta_0=0$. We thus obtain
\begin{align*}
&\lim_{\varepsilon \searrow 0} \int_{-\infty}^\infty
\frac{\cosh \beta}{1 - \cosh(2 \beta - i \varepsilon)}\: e^{2 i \ell \beta}\: d\beta
= 2 \pi \ell \sum_{n=1}^\infty \text{Res} \Big( \frac{e^{2 i \ell \beta}}{\sinh \beta} ,
\beta_n \Big) \\
&\;\;= 2 \pi \ell \sum_{n=1}^\infty \big( - e^{-2 \pi \ell} \big)^n
= - 2 \pi \ell \frac{e^{-2 \pi \ell} }{1 + e^{-2\pi\ell}}
=-\frac{2 \pi \ell}{1 + e^{2\pi\ell}} .
\end{align*}
This concludes the proof.
\QED

\section{Self-Adjoint Extension of the Relative Fermionic Signature Operator} \label{secFA}
We let~$U$ be the mapping
\beq \label{Utrans}
U : \H \rightarrow L^2(\R, \C^2)\:,\qquad g(s,\alpha) \mapsto
\hat{g}(s,\ell) = \frac{1}{\sqrt{8 \pi m}} \int_{-\infty}^\infty g(s,\alpha) \: e^{i \ell \alpha}\: d\alpha \:.
\eeq
From~\eqref{prpsprod} and Plancherel's theorem, one sees immediately that this mapping is
unitary. Moreover, its inverse is given by
\beq \label{Uinv}
U^{-1} : L^2(\R, \C^2) \rightarrow \H \:,\qquad \hat{g}(s,\ell) \mapsto g(s,\alpha)
= \sqrt{\frac{2m}{\pi}} \int_{-\infty}^\infty \hat{g}(s,\ell) \: e^{-i \ell \alpha}\: d\ell \:.
\eeq
 
\begin{Thm} \label{thmDSig}
The relative fermionic signature operator defined by~\eqref{SigRdef}
with domain~$\Cisc(\scrM, S\scrM) \cap \H \subset \H \;(\equiv \H_\scrM)$ is essentially
self-adjoint. The domain of its unique self-adjoint extension is given by
\beq \label{DSig}
\D(\Sig_\scrR) = U^{-1} \Big( \Big\{ \hat{g} \in L^2(\R, \C^2) \quad \text{with} \quad
\hat{\Sig}_\scrR \, \hat{g} \in L^2(\R, \C^2) \Big\} \Big)
\eeq
(where~$(\hat{\Sig}_\scrR \hat{g})(\ell) = \hat{\Sig}_\scrR(\ell) \, \hat{g}(\ell)$ is the
pointwise multiplication by the matrix in~\eqref{hatSdef}).
The spectrum of the relative signature operator
consists of a pure point spectrum at zero and an absolutely continuous spectrum,
\[ \sigma_\text{\rm{pp}} \big( \Sig_\scrR \big) = \{0\}\:,\qquad
\sigma_\text{\rm{ac}} \big( \Sig_\scrR \big) = \R\:. \]
It has the spectral decomposition
\[ \Sig_\scrR = \int_{-\infty}^\infty \lambda\: dE_\lambda \:, \]
where the spectral measure~$dE_\lambda$ is given by
\[ E_V = U^{-1} \Big( \chi_V(0) \,\hat{K} + \chi_V \hat{L} \Big) U \:. \]
Here~$\chi_V$ is the characteristic function of subset~$V \subset \R$,
and~$\hat{K}$ and~$\hat{L}$ are the multiplication operators
\begin{align}
\hat{L}(\ell) &= \frac{\pi m}{\ell} \: \hat{\Sig}_\scrR(\ell) = 
\begin{pmatrix} \displaystyle \frac{1}{1 + e^{-2\pi \ell}} & \displaystyle -\frac{i}{2 \cosh(\pi \ell)} \\[.9em]
\displaystyle \frac{i}{2 \cosh(\pi \ell)} & \displaystyle \frac{1}{1 + e^{2\pi \ell}}
\end{pmatrix} \\[0.3em]
\hat{K}(\ell) &= \1_{\C^2} - \hat{L}(\ell)
= \begin{pmatrix} \displaystyle \frac{e^{-2\pi \ell}}{1 + e^{-2\pi \ell}} & \displaystyle \frac{i}{2 \cosh(\pi \ell)} \\[.9em]
\displaystyle -\frac{i}{2 \cosh(\pi \ell)} & \displaystyle \frac{e^{2\pi \ell}}{1 + e^{2\pi \ell}}  \end{pmatrix} \:. \label{Ldef}
\end{align}
\end{Thm} \noindent
In order to avoid confusion, we note that the kernel of the operator~$\Sig_\scrR$ as described
by the operator~$\hat{K}$
consists of all Dirac solutions supported in the region~$\scrM \setminus \scrR$ outside the Rindler wedge.
This will be explained in detail in the proof of Theorem~\ref{thmrindler} below.

\Proof[Proof of Theorem~\ref{thmDSig}.] For a Dirac solution~$\Psi \in \Cisc(\scrM, S\scrM)$, we know from 
Lem\-ma~\ref{lemmaschwartz} and Proposition~\ref{prpsprod} that the corresponding function~$g(s,\alpha)$
is smooth and that all its derivatives are square integrable.
As a consequence, its Fourier transform is pointwise bounded and has rapid decay, i.e.
\[ \sup_{\ell} \big|(1+\ell^2)^p \:\hat{g}(\ell) \big| < \infty \qquad \text{for all~$p$} \:. \]
Using furthermore that the kernel~$I_\varepsilon(s,\alpha, \tilde{s}, .)$ given in~\eqref{Iform} decays exponentially,
we may use Fubini's theorem to exchange the orders of integration in the following computation,
\begin{align*}
\big(\Sig_\scrR & \Psi \big)(s, \alpha) = \sum_{\tilde{s}=\pm1} \lim_{\varepsilon \searrow 0} \int_{-\infty}^\infty
I_\varepsilon \big(s,\alpha; \tilde{s}, \tilde{\alpha} \big) 
\sqrt{\frac{2m}{\pi}} \bigg( \int_{-\infty}^\infty \hat{g}(s,\ell) \: e^{-i \ell \tilde{\alpha}}\: d\ell \bigg) \: d\tilde{\alpha} \\
&= \sqrt{\frac{2m}{\pi}}  \sum_{\tilde{s}=\pm1} \int_{-\infty}^\infty \hat{g}(s,\ell) \: 
\bigg( \lim_{\varepsilon \searrow 0} \int_{-\infty}^\infty
I_\varepsilon \big(s,\alpha; \tilde{s}, \tilde{\alpha} \big) \:e^{-i \ell \tilde{\alpha}}\:d\alpha \bigg) d\ell \\
&= \sqrt{\frac{2m}{\pi}}  \sum_{\tilde{s}=\pm1} \int_{-\infty}^\infty 
\big(\hat{\Sig}_\scrR(\ell) \:\hat{g}(\ell)\big)_s \: \: e^{-i \ell \alpha}\:  d\ell = \big( U^{-1} \,\hat{\Sig}_\scrR\, U \Psi \big)(s,\alpha) \:,
\end{align*}
where in the last line we applied Lemma~\ref{lemmadiagonal}. Therefore, the unitary transformation
of~$\Sig_\scrR$ yields a multiplication operator, i.e.
\[ \big( U\, \Sig_\scrR\, U^{-1} \hat{g}\big)(\ell) = \hat{\Sig}_\scrR(\ell)\, \hat{g}(\ell)
\qquad \text{for all~$\hat{g} \in U \big( \Cisc(\scrM, S\scrM) \cap \H \big)$}\:. \]
Obviously, this multiplication operator can be extended to the domain
\beq \label{Dmult}
\D \big( \,\hat{\Sig}_\scrR \big) := \Big\{ \hat{g} \in L^2(\R, \C^2) \quad \text{with} \quad
\hat{\Sig}_\scrR \, \hat{g} \in L^2(\R, \C^2) \Big\}
\eeq
(where again~$(\hat{\Sig}_\scrR \hat{g})(l) := \hat{\Sig}_\scrR(\ell) \, \hat{g}(\ell)$).
Our first task is to prove that with this domain, the multiplication operator~$\hat{\Sig}_\scrR$ is self-adjoint.
Once this has been shown, we obtain the self-adjointness of~$\Sig_\scrR$ with domain~\eqref{DSig}
by unitary transformation. Moreover, the properties of the spectrum and the spectral measure
follow immediately by computing the spectral measure of the multiplication
operator~$\hat{\Sig}_\scrR$ and unitarily transforming back to the Hilbert space~$\H$.
Our second task is to show that the domain~\eqref{DSig} coincides with the domain of the
closure of the relative fermionic signature operator.

In order to establish that the multiplication operator~$\hat{\Sig}_\scrR$ with domain~\eqref{Dmult}
is self-adjoint, we need to show that the domain of its adjoint~$\hat{\Sig}_\scrR^*$
coincides with~\eqref{Dmult}.
This follows using standard functional methods (see for example~\cite{reed+simon, lax}), which we
here recall for completeness: Let~$\Psi \in \D(\hat{\Sig}_\scrR^*)$. Then there is a vector~$\hat{\Sig}_\scrR^* \Psi
\in \H$ such that
\[ \la \Psi, \hat{\Sig}_\scrR u \ra_{L^2(\R, \C^2)} = \la \hat{\Sig}_\scrR^* \Psi, u \ra_{L^2(\R, \C^2)}
\qquad \text{for all~$u \in \D\big(\hat{\Sig}_\scrR \big)$} \:. \]
Since the function~$\hat{\Sig}_\scrR^* \Psi$ is in~$L^2(\R, \C^2)$,
we may apply Lebesgue's monotone convergence theorem to obtain
\begin{align*}
\big\| \hat{\Sig}_\scrR^* \Psi \big\|_{L^2(\R, \C^2)}
&= \lim_{L \rightarrow \infty} \big\| \chi_{[-L,L]} \,\hat{\Sig}_\scrR^* \Psi \big\|_{L^2(\R, \C^2)} \\
&= \lim_{L \rightarrow \infty} \sup_{\Phi \in \H, \; \|\Phi\|=1}
\big\la \Phi,\, \chi_{[-L,L]} \,\hat{\Sig}_\scrR^* \Psi \big\ra_{L^2(\R, \C^2)} \\
&\overset{(*)}{=} \lim_{L \rightarrow \infty} \sup_{\Phi \in \H, \; \|\Phi\|=1}
\big\la \hat{\Sig}_\scrR \,\chi_{[-L,L]} \,\Phi,\, \Psi \big\ra_{L^2(\R, \C^2)} \\
&= \lim_{L \rightarrow \infty} \sup_{\Phi \in \H, \; \|\Phi\|=1} \int_{-L}^L
\big\la \Phi(\ell),\, \hat{\Sig}_\scrR(\ell) \,\Psi(\ell) \big\ra_{\C^2}\: d\ell \\
&= \lim_{L \rightarrow \infty} \bigg( \int_{-L}^L
\big\| \hat{\Sig}_\scrR(\ell) \,\Psi(\ell) \big\|_{\C^2}^2 \: d\ell \bigg)^\frac{1}{2} \:,
\end{align*}
where in~$(*)$ we used that the function~$\chi_{[-L,L]} \,\Phi$ is in the domain of~$\hat{\Sig}_\scrR$
(see~\eqref{Dmult} and exploited the fact that the matrix~$\hat{\Sig}_\scrR(\ell)$ in~\eqref{hatSdef} is uniformly
bounded for~$\ell \in [-L,L]$).
Again applying Lebesgue's monotone convergence theorem, we infer that
the pointwise product~$\hat{\Sig}_\scrR(\ell) \,\Psi(\ell)$ is in~$L^2(\R, \C^2)$.
Using~\eqref{Dmult}, it follows that the vector~$\Psi$ lies in the domain of~$\hat{\Sig}_\scrR$.

It remains to be shown that the domain~\eqref{DSig} coincides with the domain of the
closure~$\overline{\Sig}_\scrR$ of the relative fermionic signature operator~$\Sig_\scrR$
(the latter operator with domain~$\Cisc(\scrM, S\scrM) \cap \H$).
To this end, it suffices to prove that the unitary transformation of the domain of the closure
contains all Schwartz functions,
\beq \label{USchwartz}
U \,\D ( \overline{\Sig}_\scrR ) \supset {\mathcal{S}}(\R, \C^2) \:,
\eeq
because then every vector in~\eqref{Dmult} can be approximated in the corresponding
graph norm by a sequence of Schwartz functions
(this sequence can be constructed for example by
mollification and multiplication by smooth cutoff functions).
In order to prove~\eqref{USchwartz}, let~$\hat{g}$ be a Schwartz function.
Then, taking the Fourier transforms~\eqref{Uinv} and~\eqref{fourierrep},
the resulting function~$\Psi$ is again a Schwartz function when restricted to the Cauchy surface at~$t=0$,
\[ \Psi|_{t=0} \in  {\mathcal{S}}(\R, \C^2) \:. \]
We now choose a sequence~$\Psi_n \in \Cisc(\scrM, S\scrM) \cap \H$ such that
\[ \Psi_n|_{t=0} \rightarrow \Psi|_{t=0} \qquad \text{in~${\mathcal{S}}(\R, \C^2)$} \:. \]
Taking the inverse transformation of~\eqref{fourierrep} 
(see the explicit formula~\eqref{gpmdef}) as well as the Fourier transform~\eqref{Utrans},
the resulting sequence~$\hat{g}_n$ converges to~$\hat{g}$ in~${\mathcal{S}}(\R, \C^2)$
(simply because the Fourier transform and multiplication by polynomially bounded
functions are continuous mappings on the Schwartz space).
Since the graph norm in~\eqref{Dmult} is bounded by the Schwartz norms,
the sequence~$(\hat{g}_n, \hat{\Sig}_\scrR \hat{g}_n)$ converges in the graph
norm to~$(\hat{g}, \hat{\Sig}_\scrR \hat{g})$. Hence~$\hat{g}$ indeed lies in~\eqref{USchwartz}.
This concludes the proof.
\QED

\section{The Fermionic Signature Operator of Rindler Space-Time}
Having defined the relative fermionic signature operator~$\Sig_\scrR$
as a self-adjoint operator with dense domain~$\D(\Sig_\scrR)$,
the fermionic signature operator~$\Sig$ in Rindler space-time is obtained from~\eqref{Sigintrinsic}.
We then have the following result.
\begin{Thm} \label{thmrindler} Choosing the domain of definition
\beq \label{DSigR}
\D(\Sig) = \pi_\scrR \D(\Sig_\scrR)
\eeq
(with~$\D(\Sig_\scrR)$ according to~\eqref{DSig}), the fermionic signature operator~$\Sig$
in Rindler space-time is a self-adjoint operator on~$\H_\scrR$.
It has the absolutely continuous spectrum
\beq \label{sigmaac}
\sigma_\text{\rm{ac}}(\Sig)=\R
\eeq
with spectral measure~$dE_\lambda$ given by
\[ E_V = \pi_\scrR \:U^{-1} \big( \,\chi_V \hat{L}\, \big) U\: \iota_\scrM \:, \]
where~$\hat{L}$ is again the multiplication operator~\eqref{Ldef}.
\end{Thm}
\Proof On the solution space~$\H_\scrM$ in Minkowski space, we consider the transformation
\[ \TCPT \::\: \H_\scrM \rightarrow \H_\scrM \:, \qquad \Psi(t,x) \mapsto \gamma^0 \gamma^1 \:\Psi(-t,-x) \]
(in physics referred to as the CPT transformation~\cite[Section~5.4]{bjorken};
one verifies directly that this transformation maps again to solutions of the Dirac equation). A direct computation shows that~$\TCPT$ is unitary and that~$\TCPT^2=-\1$.

The transformation~$\TCPT$ can be used to describe the Hilbert space~$\H_\scrM$ completely
in terms of~$\H_\scrR$. To see how this comes about, we first note that a solution~$\Psi \in \H_\scrM$
is determined uniquely by its Cauchy data at time zero. The restriction to the right half line~$\Psi|_{\{t=0, x>0\}}$
gives rise to a unique solution in~$\H_\scrR$, and applying~$\iota_\scrM$ yields a solution
in Minkowski space which vanishes identically on the left half line~$\Psi|_{\{t=0, x<0\}}$.
Applying~$\TCPT$ to this solution gives a new solution which vanishes identically on the
right half line~$\Psi|_{\{t=0, x>0\}}$. In view of~\eqref{printM}, the solutions which vanish on the right
half line are orthogonal to those which vanish on the left half line. We thus obtain the
orthogonal direct sum decomposition
\[ \H_\scrM = \big( \TCPT \,\iota_\scrM\, \H_\scrR \big) \oplus \big( \iota_\scrM \,\H_\scrR \big) \:. \]

Since the Dirac solutions in~$\TCPT \,\iota_\scrM \,\H_\scrR$ vanish identically in the Rindler wedge,
it is obvious that
\[ \Sig_\scrR \big|_{\TCPT \,\iota_\scrM\, \H_\scrR} = 0 \qquad \text{and} \qquad
\Sig_\scrR \big(\H_\scrM \big) \subset \iota_\scrM \H_\scrR \:. \]
Moreover, working out~$\TCPT$ in momentum space, one sees that~$\TCPT$
leaves the parameter~$\ell$ in~\eqref{planewave} unchanged and simply
maps the trivial and non-trivial eigenspaces of the matrix~\eqref{hatSdef} to each other
(see~\eqref{eigenvals} and~\eqref{eigenvectors}).
This shows that the operator~$\iota_\scrM$ in~\eqref{Sigintrinsic} maps precisely
to the orthogonal complement of the kernel of~$\Sig_\scrR$, and that the image of~$\Sig_\scrR$
is mapped by~$\pi_\scrM$ unitarily to~$\H_\scrR$.
Therefore, the spectral representation of~$\Sig$ is obtained by that of~$\Sig_\scrR$ simply
by removing the kernel. This gives the result.
\QED

As an immediate consequence, we obtain a contradiction to $m$-finiteness (see~\cite[Definition~3.3]{finite}).

\begin{Corollary} {\bf{(Rindler space-time is not $m$-finite)}} \label{corminfinite} 
For any~$c>0$ there is a vector~$\phi \in \Cisc(\scrR, S\scrR) \cap \H$ such that
\[ \big| \bra \phi | \phi \ket \big| > c\, \|\phi\|^2 \]
(where $\bra .|. \ket$ is the space-time inner product~\eqref{stip},
and~$\| . \|$ is the norm corresponding to the scalar product~\eqref{print}).
\end{Corollary}
\Proof Assume conversely that there is~$c>0$ such
that~$|\bra \phi | \phi \ket| \leq c\, \|\phi\|^2$ for all~$\phi \in \Cisc(\scrR, S\scrR) \cap \H$.
Then, using~\eqref{Sdef}, it follows that
\[ \big| (\phi \,|\, \Sig \,\phi) \big| \leq c\,  \|\phi\|^2 \qquad \text{for all~$\phi \in \Cisc(\scrR, S\scrR) \cap \H$}. \]
Since~$\Cisc(\scrR, S\scrR) \cap \H$ is dense in~$\H$, we conclude that the operator~$\Sig$ is bounded,
in contradiction to~\eqref{sigmaac}.
\QED

\section{Connection to the Hamiltonian in Rindler Coordinates}
The fermionic signature operator is closely related to the Dirac Hamiltonian in Rindler
coordinates, as we now explain.
Recall that the {\em{Rindler coordinates}}~$\tau \in \R$ and~$\rho \in (0, \infty)$
are defined by
\[ \begin{pmatrix} t \\ x \end{pmatrix} = \rho \begin{pmatrix} \sinh \tau \\ \cosh \tau \end{pmatrix} \:. \]
In these coordinates, the Rindler line element takes the form
\[ ds^2 = \rho^2\, d\tau^2 - d\rho^2 \:. \]
We work intrinsically in Rindler space-time. Translations in the time coordinate~$\tau$,
\beq \label{boost}
\tau \mapsto \tau + \Delta \:,\qquad \rho \mapsto \rho \:,
\eeq
describe a Killing symmetry. Therefore, writing the Dirac equation in this time coordinate in the Hamiltonian form
\beq \label{rindH}
i \partial_\tau \psi = H \psi \:,
\eeq
the Dirac Hamiltonian is time independent (for details see the proof of Theorem~\ref{thmFRV} below).

\begin{Thm} \label{thmFRV} The fermionic signature operator~$\Sig$ and the Hamiltonian~$H$
in Rindler coordinates satisfy the relation
\[ \Sig = -\frac{H}{\pi m} \:. \]
\end{Thm}
\Proof One method of deriving the Dirac operator would be to compute the spin connection in this coordinate
system. For our purposes, it is more convenient to again take the Dirac operator in the reference frame~$(t,x)$
and to express it in the Rindler coordinates~$(\tau, \rho)$, but without transforming the spinor basis
(this Dirac operator coincides with the intrinsic Dirac operator up to a local $\U(1,1)$-gauge transformation; for
details in the more general four-dimensional setting see~\cite{U22}). Using the identities
\begin{align*}
\frac{\partial}{\partial \rho} &= \frac{\partial t}{\partial \rho}\, \frac{\partial}{\partial t} + 
\frac{\partial x}{\partial \rho}\, \frac{\partial}{\partial x}
= \sinh \tau\: \frac{\partial}{\partial t} + 
\cosh \tau \: \frac{\partial}{\partial x} \\
\frac{\partial}{\partial \tau} &= \frac{\partial t}{\partial \tau}\, \frac{\partial}{\partial t} + 
\frac{\partial x}{\partial \tau}\, \frac{\partial}{\partial x}
= \rho \cosh \tau\: \frac{\partial}{\partial t} + 
\rho \sinh \tau \: \frac{\partial}{\partial x} \:,
\end{align*}
the Dirac operator becomes
\begin{align*}
\Dir &=  \frac{i}{\rho} \Big( \gamma^0\cosh\tau - \gamma^1 \sinh\tau \Big) \,\partial_\tau
+ i \Big( -\gamma^0 \sinh \tau + \gamma^1 \cosh \tau \Big) \,\partial_\rho \\
&= \frac{i}{\rho} \begin{pmatrix} 0 & e^{-\tau} \\ e^\tau & 0 \end{pmatrix} \,\partial_\tau
+ i \begin{pmatrix} 0 & e^{-\tau} \\ -e^{\tau} & 0 \end{pmatrix} \partial_\rho  \:.
\end{align*}
Consequently, the Dirac Hamiltonian in~\eqref{rindH} can be written as
\begin{align*}
H 
&= i \rho \begin{pmatrix} 1 & 0 \\ 0 & -1 \end{pmatrix} \partial_\rho
+ m \rho \begin{pmatrix} 0 & e^{-\tau} \\ e^{\tau} & 0 \end{pmatrix} \:.
\end{align*}

The time translation in~\eqref{boost} must be complemented by the corresponding transformation of the spinors
\beq \label{spinor}
\psi \mapsto \exp\Big( \gamma^0 \gamma^1\: \frac{\Delta}{2} \Big) \psi
= \begin{pmatrix} e^{-\frac{\Delta}{2}} & 0 \\ 0 & e^{\frac{\Delta}{2}} \end{pmatrix} \psi \:. 
\eeq
Indeed, by direct computation one verifies that the Dirac operator as well as the Dirac Hamiltonian are
invariant under the joint transformations~\eqref{boost} and~\eqref{spinor}.
If we also change the momentum variables according to
\beq \alpha \mapsto \alpha + \Delta \:, \label{momentum}
\eeq
we know by Lorentz symmetry that the Dirac solutions in our Fourier representation remain unchanged.
Therefore, the time evolution in the time coordinate~$\tau$ is described by the inverse of the transformation~\eqref{momentum}, $\alpha \mapsto \alpha - \Delta$.
We conclude that, infinitesimally, the Hamiltonian~$H$ is given by~$i \partial_\tau = i \partial_\Delta
= -i \partial_\alpha$.
Using this formula in our plane wave ansatz~\eqref{planewave}, we conclude that
\[ H \hat{g}(s,\ell) = -\ell\:\hat{g}(s,\ell) \:. \]
Comparing with~\eqref{eigenvals}, one sees that the eigenvalues of~$H$ agree up to a factor~$-\pi m$
with that of those of the relative fermionic signature operator. Taking into account that the
image of the operator~$\iota_\scrM$ in~\eqref{Sigintrinsic} coincides with the orthogonal complement of
the kernel of~$\Sig_\scrR$ (see Theorem~\ref{thmrindler}), we obtain the result.
\QED

\section{Applications: The FP State and Thermal States} \label{secquantum}
As explained in~\cite{fewster+lang, hadamard}, the fermionic signature operator can also be used to single out a distinguished fermionic quantum state, sometimes
referred to as the fermionic projector state
(or FP state). We now recall the construction and show that, in two-dimensional Rindler space-time,
this construction gives precisely the Fulling-Rindler vacuum. We again work intrinsically
in Rindler space-time. Since the Hamiltonian in the Dirac equation~\eqref{rindH} is independent of~$\tau$,
we can separate the $\tau$-dependence with a plane wave ansatz
\beq \label{sept}
\psi(\tau, \rho) = e^{-i \Omega \tau}\, \chi(\rho) \:.
\eeq
The sign of the separation constant~$\Omega$ gives a splitting of the solution space
of the Dirac equation into two subspaces. The Fulling-Rindler vacuum is the unique quantum state
corresponding to this ``frequency splitting'' in the time coordinate~$\tau$.
Next, the fermionic signature operator as defined by~\eqref{Sdef}
is a self-adjoint operator with dense domain~$\D(\Sig)$ given by~\eqref{DSigR}.
Therefore, the functional calculus gives rise to projection operators~$\chi_{(-\infty, 0)}(\Sig)$
and~$\chi_{(0, \infty)}(\Sig)$. Applying Araki's construction in~\cite{araki1970quasifree}
gives the FP state~$\omega$, being a pure quasi-free state on the
algebra generated by the smeared fields operators (for details see~\cite[Section~6]{hadamard}).
In view of Theorem~\ref{thmFRV}, the projection operators~$\chi_{(-\infty, 0)}(\Sig)$
and~$\chi_{(0, \infty)}(\Sig)$ coincide with the above frequency splitting. We thus obtain the following result:
\begin{Corollary} \label{corFRV1}
The pure quasi-free FP state~$\omega$ obtained from the fermionic signature operator
coincides with the Fulling-Rindler vacuum.
\end{Corollary} \noindent
The advantage of working with the fermionic signature operator is that
the construction is robust under perturbations of the metric. 
This connection will be discussed further in Section~\ref{secoutlook} below.

Applying Theorem~\ref{thmFRV}, one can also construct thermal states
by realizing the Dirac-Fermi distribution
(see~\cite{Unruh:1976db} or~\cite[Chapter~5]{waldqft}; note that in
our units the Boltzmann constant~$k_B=1$):

\begin{Corollary} \label{corFRV2}
Applying Araki's construction to the positive operator
\beq \label{Wbeta}
W(\beta) = \frac{1}{1+ e^{\beta m \pi \Sig}} \:,
\eeq
one obtains a thermal state of temperature~$1/\beta$. Choosing~$\beta=2 \pi$,
we get the Unruh state.
\end{Corollary}

In order to illustrate the above constructions, we note that that the two-point function
of the FP state of Corollary~\ref{corFRV1} is given by the kernel of the fermionic projector~${\mathcal{P}}(x,y)$,
as we now recall. Exactly as explained in~\cite[Section~3]{finite}, the {\em{fermionic projector}}~$P$
is introduced as the operator
\[ P = -\chi_{(-\infty, 0)}(\Sig)\, k_m \::\: C^\infty_0(\scrR, S\scrR) \rightarrow \H \:, \]
where~$\Sig$ is the fermionic signature operator,
and~$k_m$ is the {\em{causal fundamental solution}} defined as the difference of the
advanced and retarded Green's operators,
\[ k_m := \frac{1}{2 \pi i} \left( s_m^\vee - s_m^\wedge \right) \::\: C^\infty_0(\scrR, S\scrR) \rightarrow \Cisc(\scrR, S\scrR) \cap \H_\scrR\:. \]
The fermionic projector~$P$ can be represented by a distribution, referred
to as the {\em{kernel of the fermionic projector}}. Namely, just as in~\cite[Section~3.5]{finite},
one shows that there is a unique distribution~${\mathcal{P}} \in \D'(\scrR \times \scrR)$ such that 
\[ \bra \phi | P \psi \ket = {\mathcal{P}} \big( \overline{\phi} \otimes \psi \big)
\qquad \text{for all~$\phi, \psi \in C^\infty_0(\scrR, S\scrR)$}\:. \]
Now the bi-distribution~${\mathcal{P}}(x,y)$ agrees with the two-point function of the
FP state in Corollary~\ref{corFRV1} constructed from~$\chi_{(-\infty, 0)}(\Sig)$.

We finally remark that, after inserting an ultraviolet regularization, the kernel of the fermionic projector
gives rise to a {\em{causal fermion system}} (see~\cite[Section~4]{finite} or~\cite[Section~1.2]{cfs}).
The theory of causal fermion systems is an approach to describe fundamental physics
(see~\cite{pfp, cfs} or the survey papers~\cite{dice2014} or~\cite{intro}).
In this context, the kernel of the fermionic projector is used extensively in the analysis of the
causal action principle.

\section{Asymptotically Rindler Space-Times} \label{secoutlook}
The main purpose of this paper was to show that the construction of
quantum states with the fermionic signature operator gives agreement
with the frequency splitting in Rindler coordinates and the construction of thermal states.
We now give an outlook which also explains the advantages of working
with the fermionic signature operator.

For the frequency splitting in Rindler coordinates, it is essential that the coordinate~$\tau$
corresponds to a symmetry of space-time.
Therefore, the construction of the Fulling-Rindler vacuum breaks down as soon as
space-time no longer has this symmetry. The frequency splitting no longer works
even if the Rindler metric is modified by a small $\tau$-dependent perturbation,
simply because the separation ansatz~\eqref{sept} can no longer be used.
However, using the fermionic signature operator has the major benefit
that the constructions in~\cite{finite, infinite} apply to arbitrary space-times, without any symmetry assumptions.

In particular, the above construction applies to curved space-times 
involving a horizon for which the
metric tends to Rindler space-time at null infinity with a suitable decay rate.
The crucial point for the construction is to establish rapid decay estimates for the
Dirac solutions in null directions~\eqref{psidecay}. These estimates have been worked out
in a more general static setting in the thesis~\cite{treude}, where also sufficient
decay properties of the metric perturbations are specified.
In particular, Corollary~4.7.5 in this thesis immediately gives the following result:

\begin{Thm} \label{thmasy}
Let~$\scrR$ be the Rindler wedge~\eqref{scrRdef} with the static Lorentzian metric
\[ ds^2 = \big(1 + A(x) \big)^2 \big(dt^2 - dx^2 \big) \:, \]
where the metric function~$A$ has the following properties:
\begin{itemize}
\item[(i)] $\|A\|_{C^k(\R^+)} < \infty$ for all~$k \in \N$.
\item[(ii)] There are constants~$C, \alpha>0$ such that
\[ \big|A(x) \big|,\: \big|A'(x) \big| \leq \frac{C}{(1+x)^\alpha} \qquad \text{for all~$x \in \R^+$}\:. \]
\end{itemize}
Then the relation~\eqref{Sdef} uniquely defines the fermionic signature operator~$\Sig$
as an operator with dense domain~$\D(\Sig) = \Cisc(\scrR, S\scrR) \cap \H$.
\end{Thm} \noindent
We remark that the methods in~\cite{treude} could be adapted to the non-static situation.

In order to analyze the spectral properties of~$\Sig$, one could adapt the perturbative
methods as developed in~\cite{hadamard} for an external potential in Minkowski space.
We expect that for sufficiently small perturbations, the resulting FP state should again be Hadamard.
However, proving this conjecture is more difficult than the construction in~\cite[Section~5]{hadamard},
mainly because the fermionic signature operator in Rindler space-time does not have a spectral gap
separating the positive and negative spectrum. Therefore, we leave this problem as a project for
future research.

\section{Extension to Four-Dimensional Rindler Space-Time} \label{secfourdim}
We now explain how our results extend to the case of four-dimensional Rindler space-time.
Thus let~$\scrM=\R^{1,3}$ be four-dimensional Minkowski space and~$\scrR$ the subset
\[ \scrR = \big\{ (t,x,y,z) \in \R^{1,3}  \text{ with } |t| < x \big\} \:. \]
The Dirac equation in Rindler space-time is formulated as the restriction of the Dirac equation
in Minkowski space to~$\scrR$ (we use the same notation and conventions as
in~\cite{bjorken, peskin+schroeder}).
Its solutions are most easily constructed by separating the $y$-and $z$-dependence
with a plane wave ansatz,
\beq \label{yzplane}
\psi(t,x,y,z) = e^{i k_y y + i k_z z}\: \tilde{\psi}(t,x) \:,
\eeq
giving the Dirac equation in~$t$ and~$x$
\[ \big( i \gamma^0 \partial_t + i \gamma^1 \partial_x \big) \tilde{\psi}(t,x)
= \big( m + \gamma^2 k_y + \gamma^3 k_z \big) \tilde{\psi}(t,x) \:. \]
Transforming to momentum space, the solutions lie on a mass shell of mass
\beq \label{mtilde}
\tilde{m} := \sqrt{m^2 + k_y^2+k_z^2} \:.
\eeq
Thus, similar to~\eqref{fourier0}, we can make the ansatz
\[ \tilde{\psi}(q) = \int_{\R^2} \frac{d^2p}{(2 \pi)^2}\:\hat{\psi}(p)\: \delta(p^2-\tilde{m}^2)\: e^{-ipq} \:, \]
giving rise to the algebraic equation
\beq \label{algeDir}
\big( \omega \gamma^0 - k \gamma^1 \big) \hat{\psi} = \big( m + \gamma^2 k_y + \gamma^3 k_z \big) \hat{\psi}
\eeq
(where again~$p=(\omega, k)$).
This equation has a two-dimensional solution space.
In analogy to~\eqref{fdef}, we choose a basis of solutions~$\f_1, \f_2$.
In the next lemma it is shown that these spinors can be chosen to have similar properties
to those stated in Lemma~\ref{lemmaspinor} and Lemma~\ref{lemmasprod}.
\begin{Lemma} Given~$k_y$ and~$k_z$, there are spinors~$\f_a(p)$ with $a=\pm 1$ which solve the Dirac 
equation~\eqref{algeDir} and satisfy the relations
\begin{align*}
\Sl \f_a(\omega, k) \,|\, \f_b(\omega, k) \Sr &= \epsilon(\omega)\: \delta_{ab} \\
\fiberpairing{ \f_a(\omega,k)}{ \gamma^0 \,{\f_b}(-\omega,k) } &= 0 \\
\fiberpairing{ \f_a(\omega,k)}{ \gamma^0 \,{\f_b}(\omega,k) } &= \frac{|\omega|}{m}\:
\delta_{ab}\:.
\end{align*}
Moreover, in the parametrization~\eqref{parameter},
\[ \Sl \f_a(s,\alpha) \,|\, \f_b(\tilde{s}, \tilde{\alpha}) \Sr = s\:\delta_{ab}\: \frac{\tilde{m}}{m} \;\left\{
\begin{array}{cl} \cosh (\beta+ i \nu_a) & \text{if~$s=\tilde{s}$} \\
\sinh (\beta+ i \nu_a) & \text{if~$s \neq \tilde{s}$}\:,
\end{array}  \right. \]
where~$\beta$ is again given by~\eqref{betadef}, and the angle~$\nu_a \in (-\frac{\pi}{2}, \frac{\pi}{2})$ is defined by
\beq \label{nuadef}
\nu_a = \arctan \bigg( \frac{a}{m} \:\sqrt{k_y^2+k_z^2}\bigg) \:.
\eeq
\end{Lemma}
\Proof After rotating our reference frame, we can assume that~$k_z=0$ and~$k_y>0$. Then in the Dirac
representation (see for example~\cite{bjorken}), the Dirac equation~\eqref{algeDir} takes the form
\beq \label{Dirmatrix}
\begin{pmatrix} \omega-m & 0 & 0 & -k+i k_y \\
0 & \omega-m & -k-i k_y & 0 \\
0 & k-i k_y & -\omega-m & 0 \\
k+i k_y & 0 & 0 & -\omega-m \end{pmatrix} \hat{\psi} = 0 \:.
\eeq
Obviously, this matrix has two invariant subspaces: one spanned by the first and fourth spinor components,
and the other spanned by the second and third spinor components.
Choosing~$\f_1$ in the first and~$\f_{-1}$ in the second of these subspaces,
the above inner products all vanish if~$a \neq b$. In the remaining case~$a=b$, one
can restrict attention to two-spinors. In order to get back to the setting in two-dimensional
Rindler space-time, we use the identity
\[ U \begin{pmatrix} \omega-m & -k \pm i k_y\\
k \pm i k_y & -\omega-m \end{pmatrix} U = \begin{pmatrix} \omega-\tilde{m} & -k \\
k & -\omega-\tilde{m} \end{pmatrix} \:, \]
where~$U$ is the matrix
\[ U = \begin{pmatrix} \cos (\nu_a/2) & i \sin(\nu_a/2) \\
i \sin(\nu_a/2)  & \cos (\nu_a/2) \end{pmatrix}  \:. \]
Now the results follow by direct computation.
\QED

Using the result of this lemma, 
we can represent the solution in analogy to~\eqref{fourierrep} by
\[ \psi(q) = \sum_{a=\pm 1} \int_{\R^2} \frac{d^2p}{2 \pi}\:\epsilon(\omega) \: \delta(p^2-m^2)\: g_a(p)  \:\f_a(p)\: e^{-ipq} \]
with two complex-valued functions~$g_{\pm 1}$.
The subsequent analysis can be extended in a straightforward way.
In particular, the kernel~$I_\varepsilon$ in Corollary~\ref{corS} is to be replaced by the kernels
\[ I_\varepsilon^{a} \big( s,\alpha; \tilde{s}, \tilde{\alpha} \big) =
\frac{1}{4 \pi^2 \,m} \times
\left\{ \begin{array}{cl} \displaystyle \frac{s \cosh (\beta + i \nu_a)}{1 - \cosh(2 \beta + i \varepsilon s)} & \text{if~$s = \tilde{s}$}  \\[1.0em]
\displaystyle -\frac{s \sinh (\beta + i \nu_a)}{1 + \cosh (2 \beta)} & \text{if~$s \neq \tilde{s}\:.$}
\end{array} \right. \]
The residues can be computed as in Lemma~\ref{lemmadiagonal} if one transforms the integrals
in the following way,
\begin{align*}
&\int_{-\infty}^\infty \frac{\sinh (\beta+i \nu_a)}{1 + \cosh(2 \beta)} \: e^{-2 i \ell \beta}\: d\beta \\
&\;\;= \cos \nu_a \int_{-\infty}^\infty \frac{\sinh (\beta)}{1 + \cosh(2 \beta)} \: e^{-2 i \ell \beta}\: d\beta
+ i \sin \nu_a \int_{-\infty}^\infty \frac{\cosh (\beta)}{1 + \cosh(2 \beta)} \: e^{-2 i \ell \beta}\: d\beta \\
&\;\;= -\cos \nu_a \int_{-\infty}^\infty \frac{d}{d\beta} \left( \frac{1}{e^\beta + e^{-\beta}} \right) \: e^{-2 i \ell \beta}\: d\beta
+ i \sin \nu_a \int_{-\infty}^\infty \frac{1}{e^\beta + e^{-\beta}} \: e^{-2 i \ell \beta}\: d\beta \\
&\;\;= \int_{-\infty}^\infty \frac{1}{e^\beta + e^{-\beta}} \left(\cos \nu_a \frac{d}{d\beta}
+ i \sin \nu_a \right) \: e^{-2 i \ell \beta}\: d\beta \\
&\;\;= \int_{-\infty}^\infty \frac{1}{e^\beta + e^{-\beta}} \,\Big(-2 i \ell \cos \nu_a
+ i \sin \nu_a \Big) \: e^{-2 i \ell \beta}\: d\beta \:,
\end{align*}
showing that the integral is obtained from the earlier integral~\eqref{int1} if one only replaces the prefactor~$\ell$
by~$\tilde{\ell}$ given by
\beq \label{tildelldef}
\tilde{\ell}_a := \ell \cos \nu_a - \frac{\sin \nu_a}{2} \:.
\eeq
The same method also applies to the integral~\eqref{int2} and again amounts to the replacement~\eqref{tildelldef}.
We conclude that the matrix in~\eqref{hatSdef}
is to be replaced by the two matrices
\[ \hat{\Sig}^a_\scrR(\ell) = \frac{\tilde{\ell}_a}{\pi m}
\begin{pmatrix} \displaystyle \frac{1}{1 + e^{-2\pi \ell}} & \displaystyle -\frac{i}{2 \cosh(\pi \ell)} \\[.9em]
\displaystyle \frac{i}{2 \cosh(\pi \ell)} & \displaystyle \frac{1}{1 + e^{2\pi \ell}}
\end{pmatrix} \:. \]
These matrices have the eigenvalues
\[ \lambda=0 \qquad \text{and} \qquad \lambda = \frac{\tilde{\ell}_a}{\pi m}\:. \]
As a consequence, the analog of Theorem~\ref{thmFRV} is the following statement:

\begin{Thm} \label{thmFRV4d} After separating the $y$- and~$z$-dependence
by the plane wave an\-satz~\eqref{yzplane}, the fermionic signature operator~$\Sig$ and the Hamiltonian~$H$
in Rindler coordinates satisfy the relations
\beq \label{Sigres}
\Sig = -\frac{H}{\pi \tilde{m}} - \frac{1}{2 \pi \,m \tilde{m}} \; \gamma^0 \gamma^1
\Big( \gamma^2 \partial_y + \gamma^3 \partial_z \Big)
\eeq
with~$\tilde{m}$ according to~\eqref{mtilde}.
\end{Thm}
\Proof Considering again a Lorentz boost, just as in the proof of Theorem~\ref{thmFRV} we find
that~$H=-\ell$. Therefore, considering as in~\eqref{Dirmatrix} the situation that~$k_z=0$ and~$k_y>0$,
we obtain on the first and fourth spinor components that
\[ \Sig^a = -\frac{H}{\pi m}\: \cos \nu_a - \frac{\sin \nu_a}{2 \pi m} \]
with~$a=1$. Similarly, on the second and third spinor components, the same formula holds
with~$a=-1$. Using~\eqref{nuadef}, we can simplify these equations to
\[ \Sig^a = -\frac{H}{\pi \tilde{m}} - \frac{a k_y}{2 \pi m \,\tilde{m}} \:. \]
By direct computation, one verifies that the operator
\[ \gamma^0 \gamma^1 \Big( \gamma^2 \partial_y + \gamma^3 \partial_z \Big) \]
has an eigenvalue~$k_y$, and the corresponding eigenspace is the subspace
spanned by the first and fourth spinor components. Likewise, the subspace
spanned by the second and third spinor components is an eigenspace to the eigenvalue~$-k_y$.
This proves~\eqref{Sigres} for the case~$k_z=0$ and~$k_y>0$.
The general case follows immediately because the operator~\eqref{Sigres} is invariant under rotations
in the $yz$-plane.
\QED
We remark that the separation of the $y$- and $z$-dependence could be described
more mathematically by a Fourier transformation~$\psi(t,x,y,z) \mapsto \tilde{\psi}(t,x, k_y, k_z)$,
being a unitary transformation between corresponding Hilbert spaces.
Since this procedure is very similar to that at the beginning of Section~\ref{secFA},
we leave the details to the reader. Carrying out this procedure, the factors~$1/\tilde{m}$
become multiplication operators in momentum space (see~\eqref{mtilde}).
Clearly, in position space, these operators are nonlocal in the variables~$y$ and~$z$.

Applying the constructions outlined in Section~\ref{secquantum}, we again get
quasi-free quantum states. However, these states are different from the Fulling-Rindler vacuum
and the thermal states as obtained in Corollaries~\ref{corFRV1} and~\ref{corFRV2}.
The physical significance of these new states is presently under investigation.

\Thanks {{\em{Acknowledgments:}}
We would like to thank the referees for helpful suggestions.
F.F.\ is grateful to the Center of Mathematical Sciences and Applications at
Harvard University for hospitality and support.


\begin{thebibliography}{10}

\bibitem{araki1970quasifree}
H.~Araki, \emph{On quasifree states of {${\rm CAR}$} and {B}ogoliubov
  automorphisms}, Publ. Res. Inst. Math. Sci. \textbf{6} (1970/71), 385--442.

\bibitem{bjorken}
J.D. Bjorken and S.D. Drell, \emph{Relativistic {Q}uantum {M}echanics},
  McGraw-Hill Book Co., New York, 1964.

\bibitem{collini}
G.~Collini, V.~Moretti, and N.~Pinamonti, \emph{``{T}unnelling'' black-hole
  radiation with {$\varphi^3$} self-interaction: one-loop computation for
  {R}indler {K}illing horizons}, arXiv:1302.5253 [gr-qc], Lett. Math. Phys.
  \textbf{104} (2014), no.~2, 217--232.

\bibitem{dappiaggiDirac}
C.~Dappiaggi, T.-P. Hack, and N.~Pinamonti, \emph{The extended algebra of
  observables for {D}irac fields and the trace anomaly of their stress-energy
  tensor}, arXiv:0904.0612 [math-ph], Rev. Math. Phys. \textbf{21} (2009),
  no.~10, 1241--1312.

\bibitem{drago+murro}
N.~Drago and S.~Murro, \emph{A new class of fermionic projectors: {M}{\o}ller
  operators and mass oscillation properties}, arXiv:1607.02909 [math-ph]
  (2016).

\bibitem{fewster+lang}
C.J. Fewster and B.~Lang, \emph{Pure quasifree states of the {D}irac field from
  the fermionic projector}, arXiv:1408.1645 [math-ph], Class. Quantum Grav.
  \textbf{32} (2015), no.~9, 095001, 30.

\bibitem{fewster2013necessity}
C.J. Fewster and R.~Verch, \emph{The necessity of the {H}adamard condition},
  arXiv:1307.5242 [gr-qc], Class. Quantum Grav. \textbf{30} (2013), no.~23,
  235027, 20.

\bibitem{sea}
F.~Finster, \emph{Definition of the {D}irac sea in the presence of external
  fields}, arXiv:hep-th/9705006, Adv. Theor. Math. Phys. \textbf{2} (1998),
  no.~5, 963--985.

\bibitem{U22}
\bysame, \emph{Local {$\rm U(2,2)$} symmetry in relativistic quantum
  mechanics}, arXiv:hep-th/9703083, J. Math. Phys. \textbf{39} (1998), no.~12,
  6276--6290.

\bibitem{pfp}
\bysame, \emph{The {P}rinciple of the {F}ermionic {P}rojector}, hep-th/0001048,
  hep-th/0202059, hep-th/0210121, AMS/IP Studies in Advanced Mathematics,
  vol.~35, American Mathematical Society, Providence, RI, 2006.

\bibitem{cfs}
\bysame, \emph{The {C}ontinuum {L}imit of {C}ausal {F}ermion {S}ystems},
  arXiv:1605.04742 [math-ph], Fundamental Theories of Physics, vol. 186,
  Springer, 2016.

\bibitem{dice2014}
F.~Finster and J.~Kleiner, \emph{Causal fermion systems as a candidate for a
  unified physical theory}, arXiv:1502.03587 [math-ph], J. Phys.: Conf. Ser.
  \textbf{626} (2015), 012020.

\bibitem{intro}
F.~Finster, J.~Kleiner, and J.-H. Treude, \emph{An {I}ntroduction to the
  {F}ermionic {P}rojector and {C}ausal {F}ermion {S}ystems}, in preparation.

\bibitem{drum}
F.~Finster and O.~M\"uller, \emph{Lorentzian spectral geometry for globally
  hyperbolic surfaces}, arXiv:1411.3578 [math-ph], Adv. Theor. Math. Phys.
  \textbf{20} (2016), no.~4, 751--820.

\bibitem{hadamard}
F.~Finster, S.~Murro, and C.~R\"oken, \emph{The fermionic projector in a
  time-dependent external potential: Mass oscillation property and {H}adamard
  states}, arXiv:1501.05522 [math-ph], J. Math. Phys. \textbf{57} (2016),
  072303.

\bibitem{finite}
F.~Finster and M.~Reintjes, \emph{A non-perturbative construction of the
  fermionic projector on globally hyperbolic manifolds {I} -- {S}pace-times of
  finite lifetime}, arXiv:1301.5420 [math-ph], Adv. Theor. Math. Phys.
  \textbf{19} (2015), no.~4, 761--803.

\bibitem{infinite}
\bysame, \emph{A non-perturbative construction of the fermionic projector on
  globally hyperbolic manifolds {II} -- {S}pace-times of infinite lifetime},
  arXiv:1312.7209 [math-ph], Adv. Theor. Math. Phys. \textbf{20} (2016), no.~5,
  1007--1048.

\bibitem{planewave}
\bysame, \emph{The fermionic signature operator and {H}adamard states in the
  presence of a plane electromagnetic wave}, arXiv:1609.04516 [math-ph], Ann.
  Henri Poincar{\'e} \textbf{18} (2017), 1671--1701.

\bibitem{Fulling:1972md}
S.A. Fulling, \emph{{Nonuniqueness of canonical field quantization in
  Riemannian space-time}}, Phys. Rev. \textbf{D7} (1973), 2850--2862.

\bibitem{hormanderhyp}
L.~H{\"o}rmander, \emph{Lectures on {N}onlinear {H}yperbolic {D}ifferential
  {E}quations}, Math\'ematiques \& Applications (Berlin) [Mathematics \&
  Applications], vol.~26, Springer-Verlag, Berlin, 1997.

\bibitem{johnston}
S.~Johnston, \emph{Feynman propagator for a free scalar field on a causal set},
  arXiv:0909.0944 [hep-th], Phys. Rev. Lett. \textbf{103} (2009), 180401.

\bibitem{lax}
P.D. Lax, \emph{Functional {A}nalysis}, Pure and Applied Mathematics (New
  York), Wiley-Interscience [John Wiley \& Sons], New York, 2002.

\bibitem{peskin+schroeder}
M.E. Peskin and D.V. Schroeder, \emph{An {I}ntroduction to {Q}uantum {F}ield
  {T}heory}, Addison-Wesley Publishing Company Advanced Book Program, Reading,
  MA, 1995.

\bibitem{radzikowski}
M.J. Radzikowski, \emph{Micro-local approach to the {H}adamard condition in
  quantum field theory on curved space-time}, Comm. Math. Phys. \textbf{179}
  (1996), no.~3, 529--553.

\bibitem{reed+simon}
M.~Reed and B.~Simon, \emph{Methods of {M}odern {M}athematical {P}hysics. {I},
  {F}unctional analysis}, second ed., Academic Press Inc., New York, 1980.

\bibitem{rejzner}
K.~Rejzner, \emph{Perturbative {A}lgebraic {Q}uantum {F}ield {T}heory}, Math.
  Phys. Stud., Springer, 2016.

\bibitem{rindler1966}
W.~Rindler, \emph{Kruskal space and the uniformly accelerated frame}, Am. J.
  Phys. \textbf{34} (1966), 1174.

\bibitem{sahlmann2001microlocal}
H.~Sahlmann and R.~Verch, \emph{Microlocal spectrum condition and {H}adamard
  form for vector-valued quantum fields in curved spacetime},
  arXiv:math-ph/0008029, Rev. Math. Phys. \textbf{13} (2001), no.~10,
  1203--1246.

\bibitem{sorkin2}
R.D. Sorkin, \emph{Scalar field theory on a causal set in histories form},
  arXiv:1107.0698 [gr-qc], J. Phys. Conf. Ser. \textbf{306} (2011), 012017.

\bibitem{treude}
J.-H. Treude, \emph{Estimates of {M}assive {D}irac {W}ave {F}unctions near
  {N}ull {I}nfinity}, Dissertation, Universit\"at Regensburg,
  http://epub.uni-regensburg.de/32344/, 2015.

\bibitem{Unruh:1976db}
W.G. Unruh, \emph{{Notes on black hole evaporation}}, Phys. Rev. \textbf{D14}
  (1976), 870.

\bibitem{wald}
R.M. Wald, \emph{General {R}elativity}, University of Chicago Press, Chicago,
  IL, 1984.

\bibitem{waldqft}
\bysame, \emph{Quantum {F}ield {T}heory in {C}urved {S}pacetime and {B}lack
  {H}ole {T}hermodynamics}, Chicago Lectures in Physics, University of Chicago
  Press, Chicago, IL, 1994.

\end{thebibliography}
\providecommand{\bysame}{\leavevmode\hbox to3em{\hrulefill}\thinspace}
\providecommand{\MR}{\relax\ifhmode\unskip\space\fi MR }
\providecommand{\MRhref}[2]{%
  \href{http://www.ams.org/mathscinet-getitem?mr=#1}{#2}
}
\providecommand{\href}[2]{#2}

\end{document}